\documentclass[a4paper,twoside,10pt]{article}

\usepackage{amssymb}

\usepackage{graphicx}
\usepackage{subfigure}

\begin{document}





\title{Emergent Spiking in Non-Ideal Memristor Networks}


\author{Ella Gale, Ben de Lacy Costello and Andrew Adamatzky}




\maketitle

\begin{abstract}
Memristors have uses as artificial synapses and perform well in this role in simulations with artificial spiking neurons. Our experiments show that memristor networks natively spike and can exhibit emergent oscillations and bursting spikes. Networks of near-ideal memristors exhibit behaviour similar to a single memristor and combine in circuits like resistors do. Spiking is more likely when filamentary memristors are used or the circuits have a higher degree of compositional complexity (i.e. a larger number of anti-series or anti-parallel interactions). 3-memristor circuits with the same memristor polarity (low compositional complexity) are stabilised and do not show spiking behaviour. 3-memristor circuits with anti-series and/or anti-parallel compositions show richer and more complex dynamics than 2-memristor spiking circuits. We show that the complexity of these dynamics can be quantified by calculating (using partial auto-correlation functions) the minimum order auto-regression function that could fit it. 
We propose that these oscillations and spikes may be similar phenomena to brainwaves and neural spike trains and suggest that these behaviours can be used to perform neuromorphic computation.
\end{abstract}



\section{Introduction}

Memristors are non-linear resistors that possess a memory~\cite{14}. They were first predicted to exist in 1971~\cite{14} and were formally discovered in device form in 2008~\cite{15} although other memristor devices had been fabricated before (for a recent review see~\cite{RevMemReRAM}). After the first paper to relate memristor theory to a real world device~\cite{15}, excitement abounded over their use for computer memory (alongside ReRAM) and neuromorphic computing, as well as suggestions that they might be more energy efficient and resilient. Neurons are believed to be both the seat of memory and the processor of the brain. Memristors, by combining memory with processing, offer a tantalising glimpse of devices which could do the same. 

Neuromorphic computing is the concept of using computer components to mimic biological neural architectures, primarily the mammalian brain. The relation between memristors and neuromorphic computing dates back to 1976 when Chua and Kang expanded the idea of the memristor to a memristive system (which has two state variables rather than the one the memristor possesses) and suggested that the Hodgkin-Huxley model of the nerve axon could be improved by incorporating memristors in place of the non-linear time dependent resistors~\cite{84}: an idea that wasn't theoretically demonstrated until 2012~\cite{247,248}. 

In general, the memristor community has concentrated on synapses rather than axons: using memristors as synapses~\cite{41,217}, combining memristors and spiking neurons in simulations (see for example~\cite{278}) and even whether synapses are memristive~\cite{239} (not as unlikely as it seems now that biological materials such as flowing blood~\cite{147}, sweat ducts~\cite{276} and slime mould~\cite{277} have been shown to be memristive). Simulations have shown that memristive connections could be used to reproduce spike-time dependent plasticity~\cite{239} (the process by which synapses adjust their connection weight to implement Hebbian learning~\cite{85}). Experiments have modelled neural action with memristors~\cite{45} and investigated the interactions between living neural cells and memristors~\cite{Ne0}.

As has been shown elsewhere, our memristors~\cite{243,SpcJ} and other memristive devices~\cite{123,macro1,261} present a `transient' current spike in response to a d.c. voltage (which we have also referred to as the `short-term memory of the memristor'). In this paper, we demonstrate that these spikes interact within small memristor networks and give rise to emergent brain-like dynamics.

When assembling multi-memristor systems in the laboratory, it is sensible to first ask which circuits are being designed by theorists and tested by simulationists for use with memristors, and overwhelmingly they investigate the Chua circuit. The original Chua circuit~\cite{256} was created to demonstrate that chaos was a real phenomena (not merely due to rounding errors in the computer simulations) and it has been suggested~\cite{248} that neurons are poised at the edge of chaos, so, it is worth investigating chaotic dynamics (and the related field of complexity) if we are interested in making circuits for neuromorphic computing. 

There have been a plethora of different versions of and alterations to the Chua circuit, as summarized in~\cite{250}, but the simplest version built~\cite{257} consists of one inductor, one resistor, two capacitors and a component called Chua's diode: a non-linear circuit element usually fabricated from several other circuit components including op amps. Itoh and Chua were the first to replace Chua's diode with a memristor~\cite{251}; they worked with the concept of an active memristor (a memristor is a passive device, but a circuit of a negative resistance and memristor can be viewed as an active memristor -- a concept which fits well with biological memristors and has been used to model them~\cite{277}). There have been many papers since detailing the rich behaviour and chaotic properties of Chua circuits containing memristors (eg. ~\cite{82},~\cite{61},~\cite{70} and ~\cite{232}). These contain simulations which use Chua's equations~\cite{14} for the perfect theoretical memristor and electronic 
experiments which replace the memristor with a circuit equivalent, presumably due to the difficulty in obtaining an actual memristor to use. 

An important step forward in the direction of real world functionality was Buscarino's paper~\cite{252} where Chua's diode in Chua's circuit was modelled using Strukov et al's phenomenological model~\cite{15} which is based on real world measurables and relates to a real memristor. The resulting simulation demonstrated chaotic behaviour~\cite{252}. This paper used a pair of Strukov memristors~\cite{15} connected in anti-parallel to give a symmetrical $I-V$ curve as a replacement for Chua's diode. They then used a voltage frequency that took the memristor to its limits (i.e. maximum and minimum resistance) to introduce asymmetry and richer behaviour. However, from this data it is not known whether the chaotic behaviour they observed in their simulations arose from the memristors themselves or from the interactions of the errors in the model, which (even with windowing functions) is weakest at the edges of the memristor. 

A recent experimental result of a possible neuromorphic building block was the Hewlett-Packard (HP) `neuristor': a circuit consisting of two memristors and two capacitors (and a load resistor) which gives `brainwave'-like dynamics from a constant voltage source~\cite{272}. This circuit also positioned the memristors in anti-parallel. 

Another area of interest is how few components a chaotic circuit can be made with. A recent paper~\cite{40} suggested that the simplest circuit capable of producing chaos could be made with three components: a capacitor, an inductor and a memristor. Thus, circuits involving memristors, capacitors and inductors look likely to product interesting dynamics. 

Furthermore, according to Chua~\cite{14} the linear combination of memristors in a circuit with only one input and one output to that circuit is indistinguishable from a memristor with a memristance value calculable by standard series and parallel resistor adding rules (`A 1-port containing only memristors is equivalent to a memristor'~\cite{14}), i.e. the memristors combine in series and in parallel similarly to resistors, which would suggest that a circuit made up of only memristors would be a trivial circuit with the same behaviour as a single memristor.

However, due to our observation of the memristor spikes, we decided to test whether circuits consisting of only real-world memristors would give rise to rich behaviour, rendering the complications of including capacitors or op amps unnecessary. Therefore we investigated interacting memristors using titanium dioxide sol-gel memristors~\cite{M1}. By using experimental prototype devices we are able to make use of the memristor's actual behaviour, whereas theoretical models of the memristor can be less useful in this regard. 

It was thought that the memristors would spike with the change of voltage and this would cause a change in resistance within a single memristor, which, with this circuit set-up would lead to a voltage change across the other memristors and thus further spikes (this idea was tested in a music-composing memristor network simulation~\cite{Mu0}). 

In this paper we investigate how binary and tertiary combinations of memristors interact and test the assertion that memristor circuits addressed only by their joint one port entry (ie there is one wire coming out and going in to that part of the circuit) are indistinguishable from a single memristor.

\subsection{Methodology}

\subsubsection{Memristor Circuits Tested}

The tested circuits are presented in table~1. From the literature and our own intuition, we expected that two memristors in anti-parallel configuration would be the most likely 2-memristor circuit to exhibit rich dynamics, see circuit no. 4 in table~1. A dynamical system can exhibit chaotic behaviour if it has at least three state variables. For this reason we decided to create circuits with three memristors, which gives us the following three separate state variables: the current through the circuit, and the voltage across two of the memristors (the third being determined by the other two in a system kept at a constant voltage). In order to maximise the anti-parallel interactions of the circuit, the memristors were wired up, two in anti-series, with one in parallel to the two in series as shown in figure~1 circuit 7. For this reason, we decided to count the number of anti-parallel interactions in the 3-memristor circuits and also take note of memristors wired up with opposite polarity in a series circuit (
anti-series), which is circuit 2 in table~1. We expect that circuit number 7 one anti-parallel and one anti-series will give the richest behaviour as it has the most anti-polarity interactions and they are of different types introducing another type of variation into the system.

\begin{table}
 \begin{tabular}{|c|c|c|}
\hline
Experiment		& Wiring Diagram	& Number and Type \\
 Number			& 			& of anti-polarity interactions	\\
\hline
 & & \\
  1	& \includegraphics[width=0.15\textwidth]{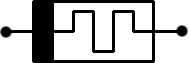} & 0\\
 & & \\
\hline
 & & \\
  2	& \includegraphics[width=0.30\textwidth]{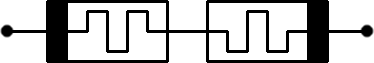}	& 1s	\\
 & & \\
\hline
 & & \\
  3	& \includegraphics[width=0.30\textwidth]{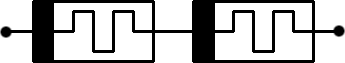}	& 0	\\
 & & \\
\hline
 & & \\
  4 	& \includegraphics[width=0.20\textwidth]{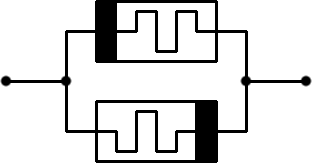}	& 1p	\\
 & & \\
\hline
 & & \\
  5 	& \includegraphics[width=0.20\textwidth]{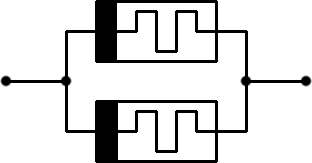}	& 0	\\
 & & \\
\hline
 & & \\
  6	& \includegraphics[width=0.30\textwidth]{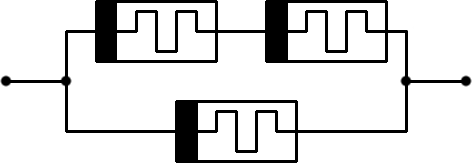}		& 0	\\
 & & \\
\hline
  & & \\
  7 	& \includegraphics[width=0.30\textwidth]{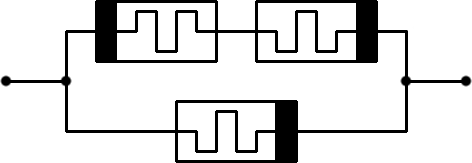}	& 1s, 1p\\
 & & \\
\hline
 & & \\
  8	& \includegraphics[width=0.30\textwidth]{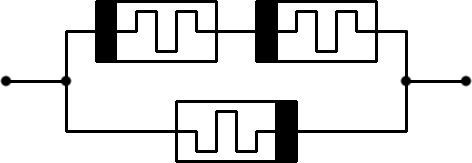}	& 2p	\\
 & & \\
\hline
 \end{tabular}
\label{tab:TestCircuits}
\caption{Constructed experiments. Anti-parallel memristor interactions are represented by `p', anti-polarity series memristor interactions by `s'.}
\end{table}

\clearpage

\subsubsection{Memristor Types: Curved and Triangular Switching Behaviours}

\begin{figure}[!tbp]
\centering
\subfigure[]{\includegraphics[width=0.49\textwidth]{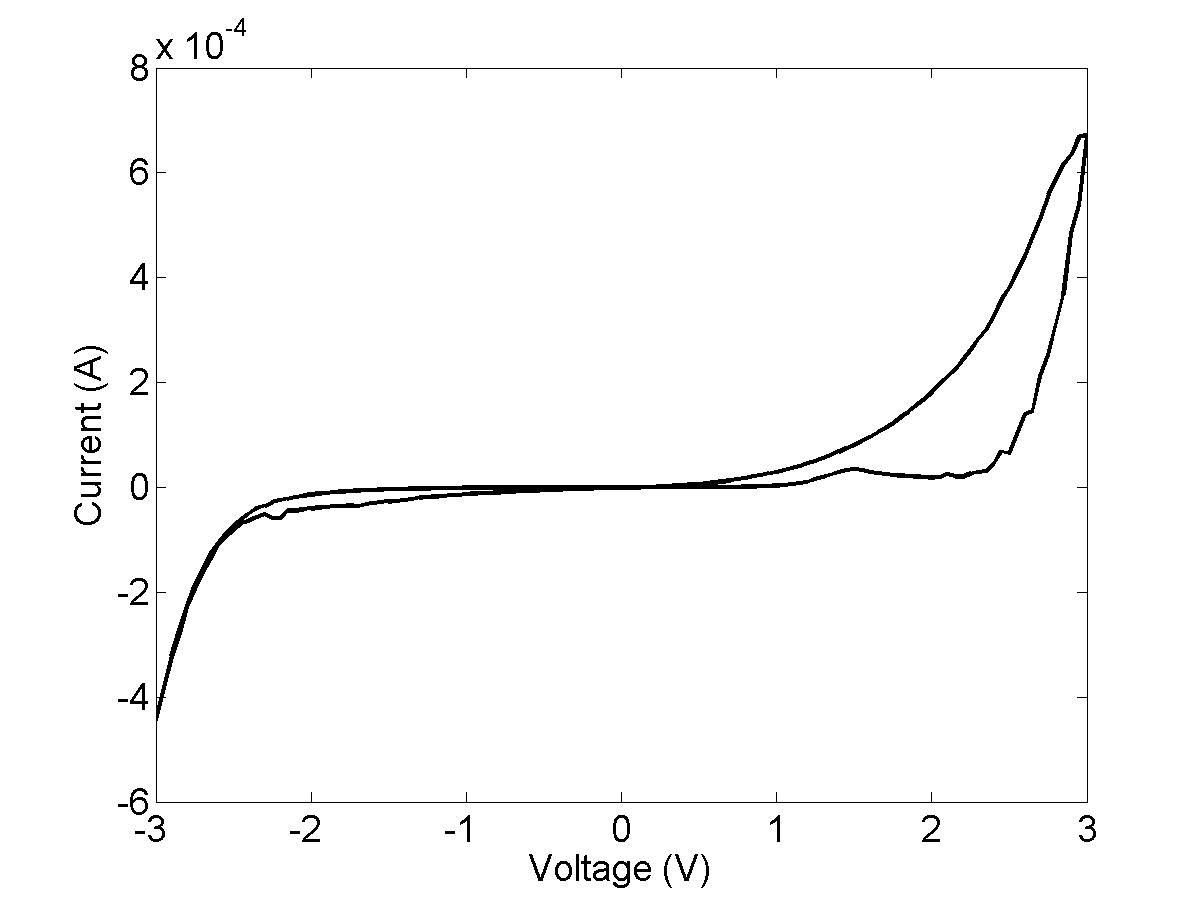}}
\subfigure[]{\includegraphics[width=0.49\textwidth]{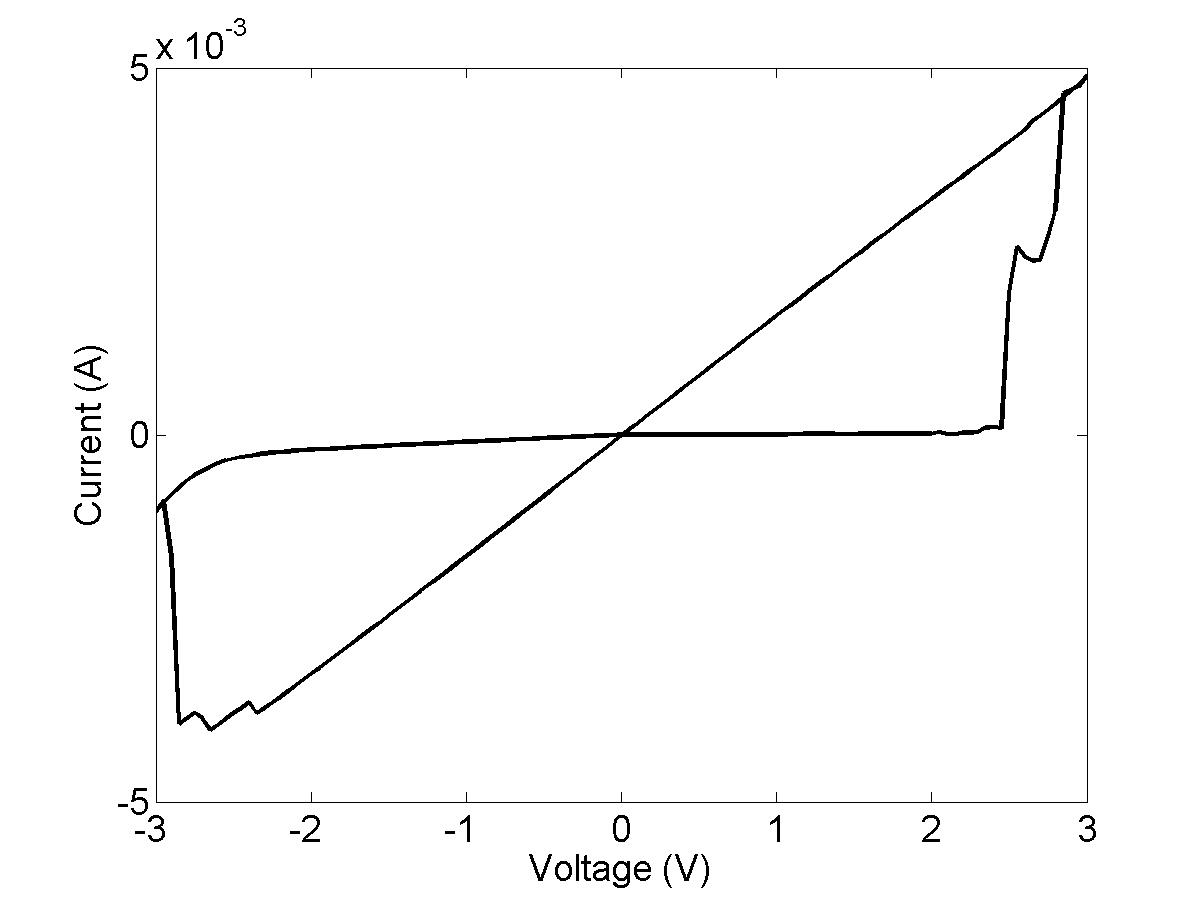}}
\subfigure[]{\includegraphics[width=0.49\textwidth]{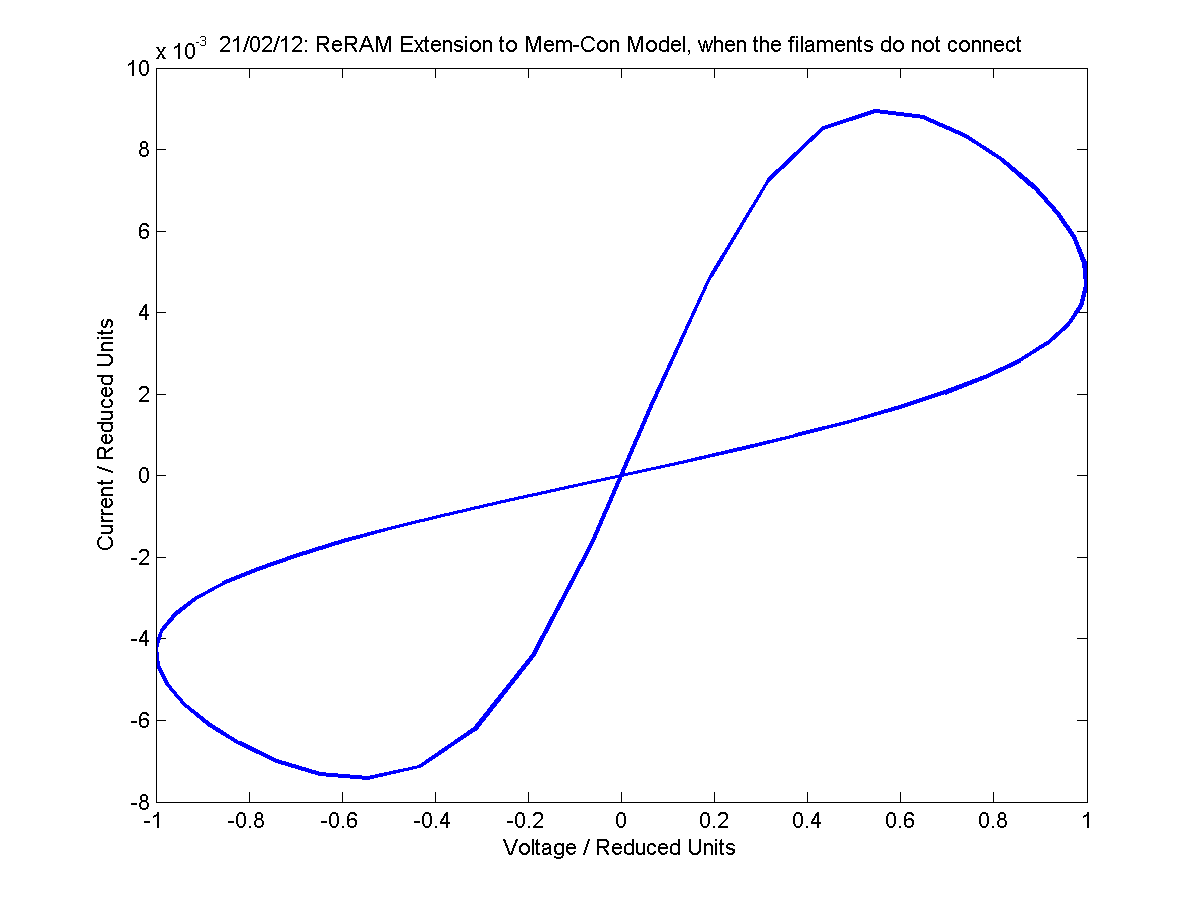}}
\subfigure[]{\includegraphics[width=0.49\textwidth]{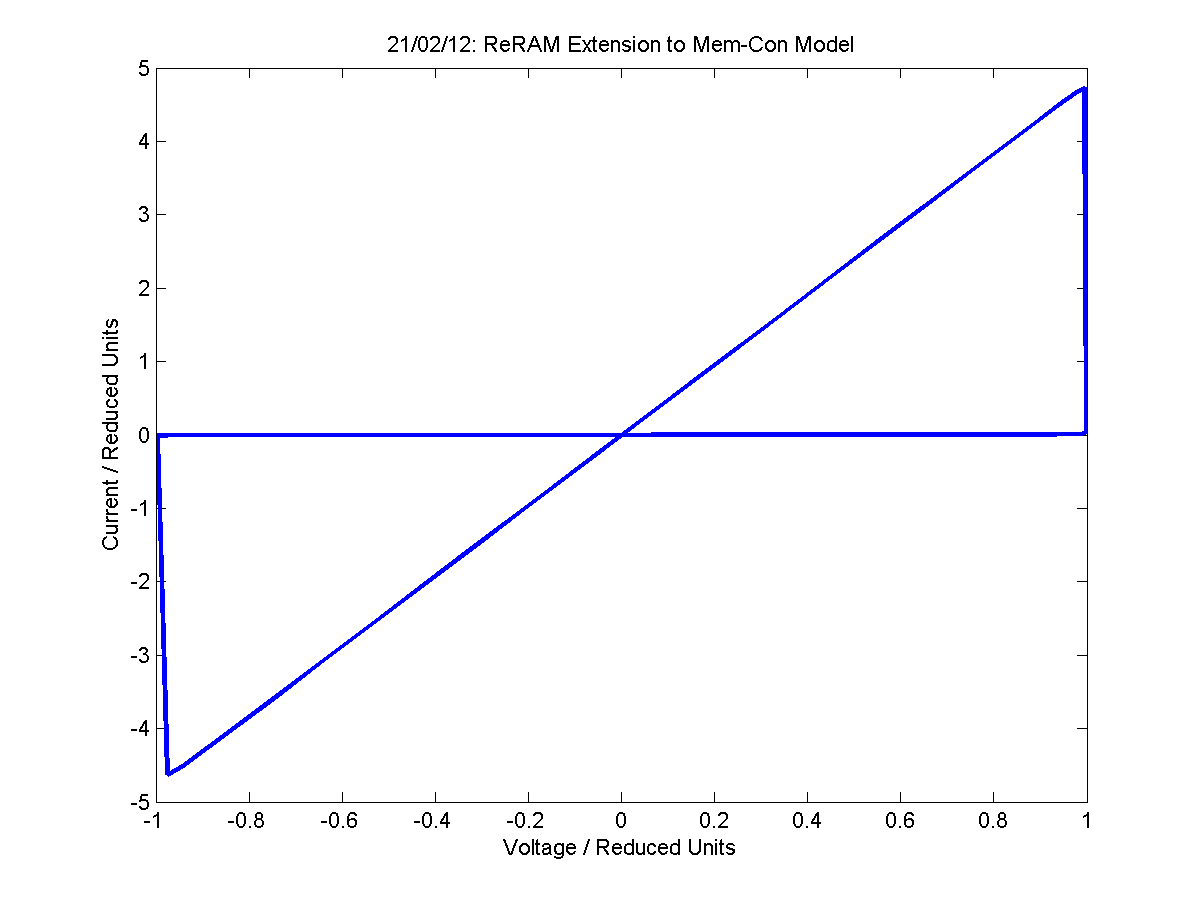}}
\caption{Examples of the two different types of memristors: a. experimental curved memristor, b. experimental filamentary memristor, c. theoretical curved type memristor, d. theoretical filamentary memristor.}
\label{graphs}
\end{figure}

Our memristors show two characteristic behaviours: A: `curved' pinched hysteresis curves and B: `triangular' pinched hysteresis curves, as previously observed~\cite{M0} and shown in figure~\ref{graphs}. The type A memristors can be modelled by the standard memory-conservation theory~\cite{F0c} and are thus close to Chua's theoretical memristor~\cite{14}. Type B memristors have an ohmic low resistance state as evidenced by a straight-line on the $I$-$V$ graph. We suspect that the `triangular' type B memristors switch via the formation and breaking of conducting filaments and the addition of a conducting filament to the memory-conservation theory of memristance~\cite{254} models this situation well. The memristors used in this work were classified into A or B types based on observed $I$-$V$ curves. All further results presented in this paper are experimental, not simulated.

\subsubsection{Experimental Details}

The circuits were wired up according to the table and connected to a Keithley 2400 Sourcemeter in current-sensing, voltage sourcing mode. To get the $I$-$t$ curves, the memristor circuits were taken to +0.4V for 1000 timesteps (1 timestep = 1.6s) the voltage source was then switched to 0V and data was gathered for a further 100 timesteps. 22 different experiments were analysed. To investigate whether a slow changing voltage had an effect, a sinusoidal voltage of 1600 timesteps of 2s was used. In all experiments voltages were kept very low to avoid the creation of filaments via Joule heating which would lead to filamentary memristors switching into lower resistance states.

The $I$-$t$ time-series plots were analysed using MatLab. The periodograms were calculated using discrete Fourier transforms using a fast-Fourier transform algorithm~\cite{DFT}: all periodograms were calculated with the same sample frequency of 0.942Hz. Time-series auto-correlation function (ACF) plots were plotted to test for significant persistence as measured by the number of timesteps (lag) required to predict the next step. Partial auto-correlation function (PACF) gives a measure of the signals `complexity' by measuring the minimum order of auto-regression (AR) function required to fit it, where an AR(0) is a memoryless system, AR(1) requires 1 preceding step and so on. Auto-correlation was performed using MatLab's `autocorr' function~\cite{Box,BoxBook} and the partial auto-correlation using MatLab's `parcorr' function~\cite{BoxBook}. Output $I$-$t$ plots were tested for a departure from randomness based on the ACF data using the Ljung-Box-Pierce Q-test~\cite{BoxBook}, implemented in MatLab as `lbqtest',
 using a threshold p-value of 0.05 (below which we reject the null hypothesis of the data being random). The test was performed lags from 1 to 50 based on the observation that bursting spikes did not tend to persist for more than 100s. 

\section{Results}

\subsection{Classification of Dynamical Response}

The 22 experiments are summarised in table~\ref{tab:2} and examples of the behaviour are shown in figures~\ref{fig:181012MTD7p2t}, ~\ref{fig:181012MTD6p1}, ~\ref{fig:241012MTR5p2}, ~\ref{fig:101012MT1p1}, ~\ref{fig:151012MT1p5} and ~\ref{fig:221012MTR6p3}. Six types of behaviour were observed and will be described in rough order of increased `complexity', where we use the term to mean a qualitative measure of the visual complexity observable in the graphs. 

\subsubsection{`Ideal' Dynamics}
The simplest dynamics were those which were single memristor-like, and showed a current transient decaying with time, see figure~\ref{fig:181012MTD7p2t} and compare to the single memristor response in figure~\ref{fig:single_mem_new}. This type of response is given the classification `1' in table 2, and accounted for 2/22 experiments. Single memristor-like $I$-$t$ curves with a few single spikes was designated `1s' (4/22), and an example is shown in figure~\ref{fig:Curved2}a. Graphs which looked like a single memristor switching state were designated `w' (2/22), see figure~\ref{fig:181012MTD6p1}. `1', `1s' and `w' were all variations on the single memristor $I$-$t$ in that they have a switching spike when the voltage was turned on or off and decay with time either without spikes (`1'), with a few spikes (`1s') or with several switching spikes (`w'). 

\subsubsection{Emergent Spiking Dynamics}
The more complex behaviour is the spiking behaviour and this is broken down into two types: `SpO' which is oscillations with a few single spikes overlaid (5/22) and `Sp' which is oscillations with bursting spikes (6/22). Examples of `SpO' responses are shown in figures~\ref{fig:151012MT1p5} and~\ref{fig:241012MTR5p2}. An example of `Sp' type dynamics are shown in~\ref{fig:101012MT1p1}. Neither the `Sp' or `SpO' circuits showed switching spikes when the voltage was turned on or off, which we think is because that spike is the impulse to the circuit and the energy emerges at a later time in the observed spikes. 

The remaining type is designated as `fi' as shown in figure~\ref{fig:Triangular2}, where the current is several orders of magnitude higher and the response more linear, both facts suggest that a conducting filament in one (or more) of the memristors has nearly bridged the electrodes (the current is linear when it finally connects). These circuits are not interesting from the dynamical point of view, however the memristors in this state are excellently suited for the task of holding a state and are useful for resistive Random Access Memory (ReRAM).

\subsection{Two Memristor Circuits}

\subsubsection{Type A (curved) Memristors in 2-Memristor Circuits}

\begin{figure}[!tbp]
\centering
\subfigure[]{\includegraphics[width=0.49\textwidth]{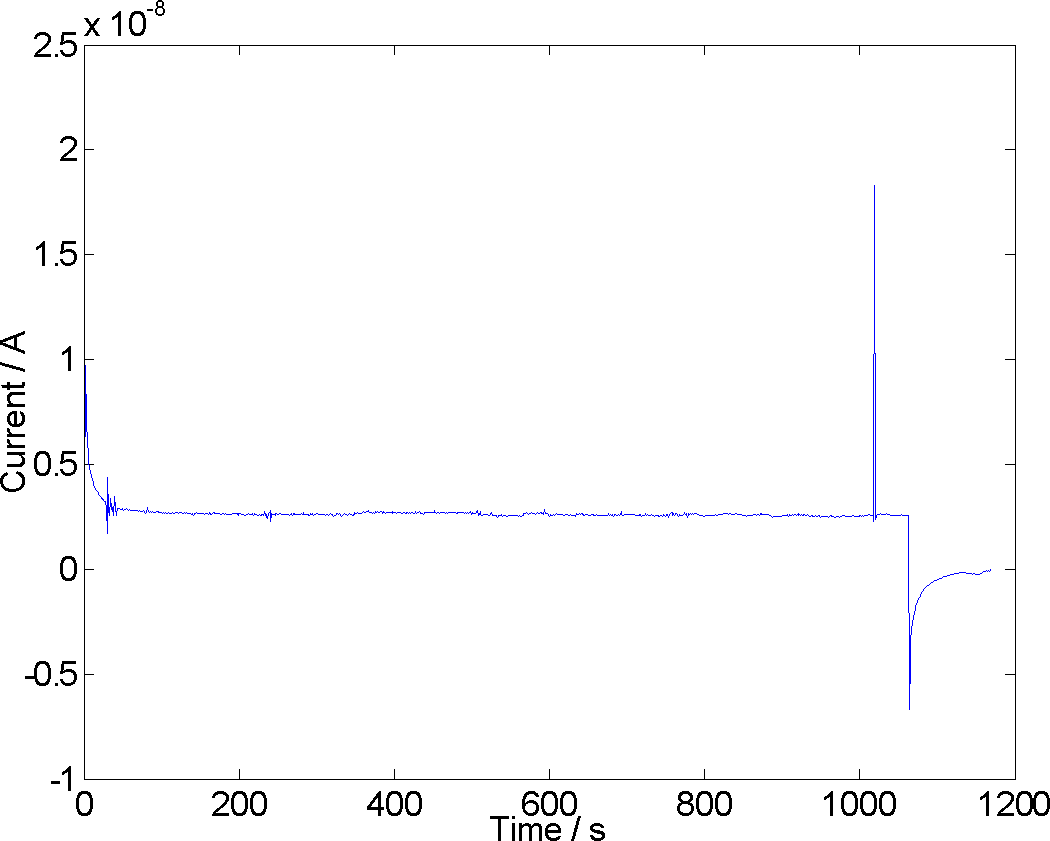}}
\subfigure[]{\includegraphics[width=0.49\textwidth]{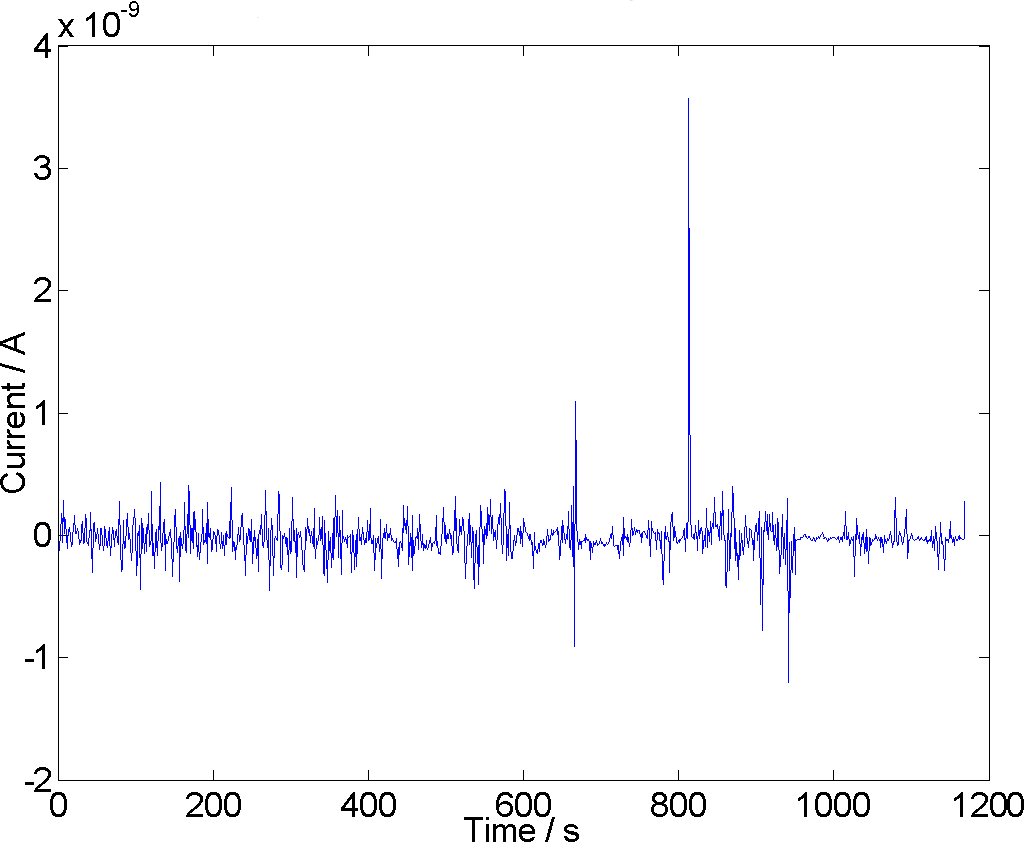}}
\subfigure[]{\includegraphics[width=0.49\textwidth]{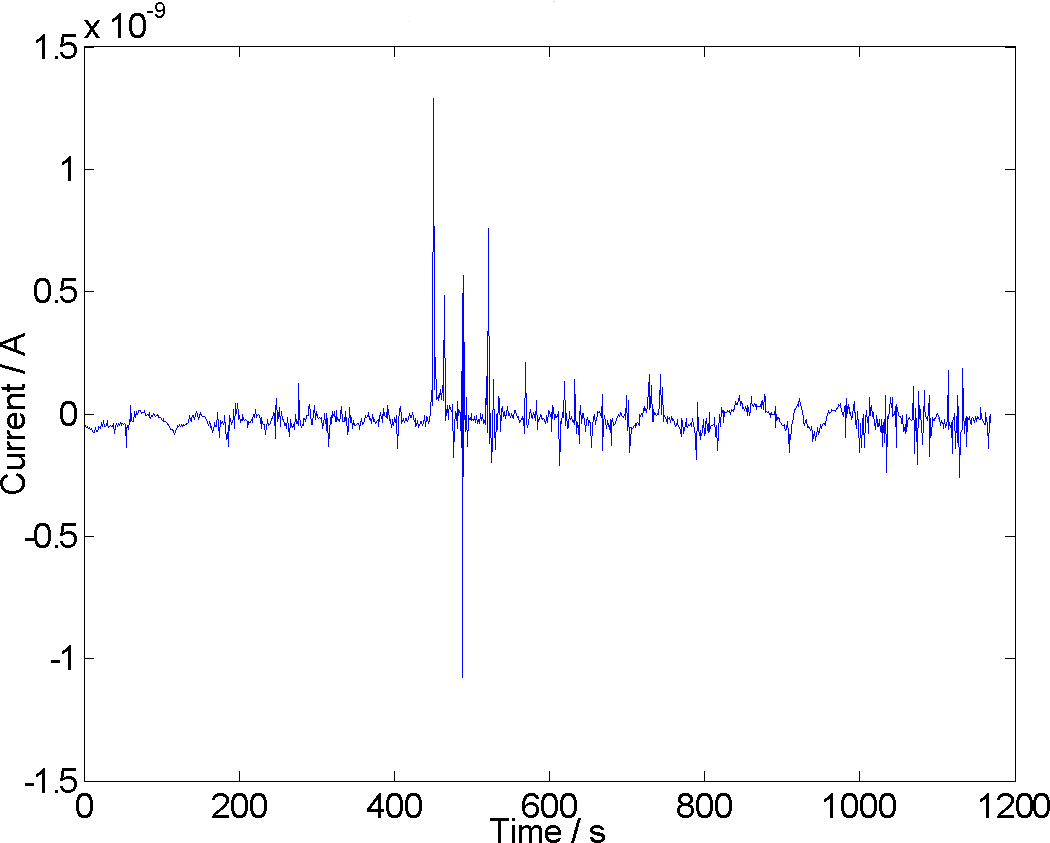}}
\subfigure[]{\includegraphics[width=0.49\textwidth]{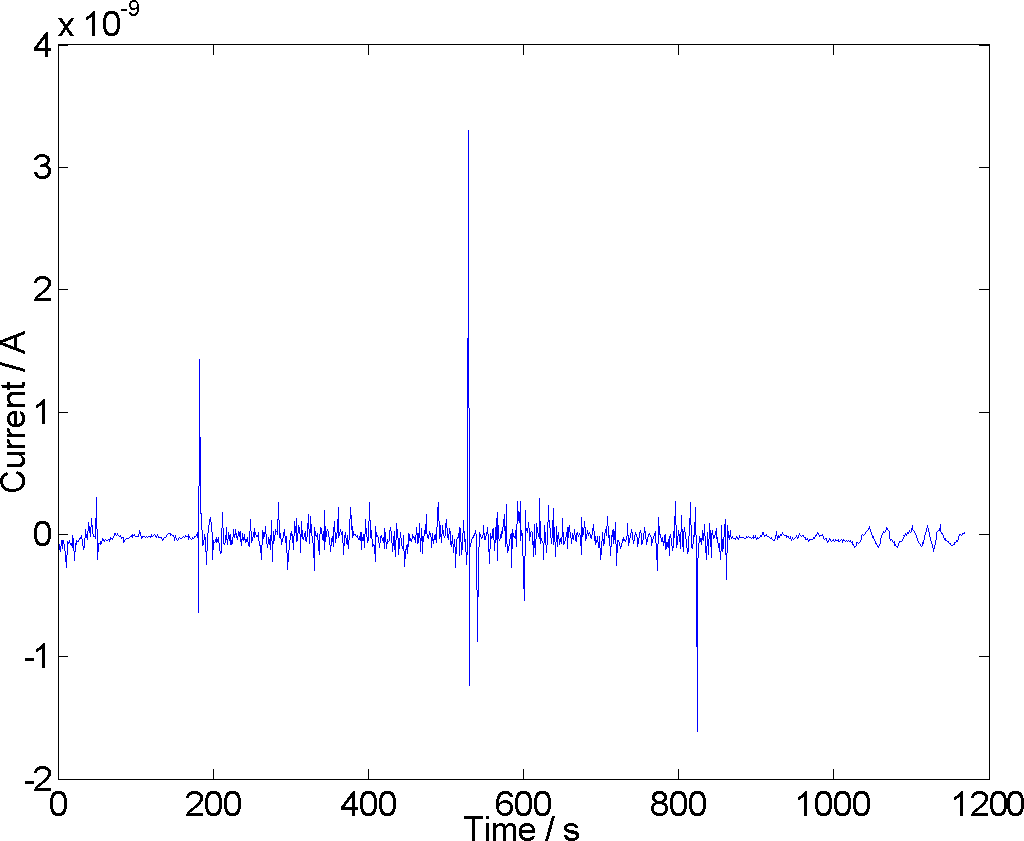}}
\caption{Results for type A memristors: a. Two A memristors in series, opposite direction, circuit 2 in table~1, b. Two type A memristors in series, same direction, circuit 3 in table~1, c. Two type A memristors in anti-parallel, circuit 4 in table~1 and d. two type A memristors in parallel, circuit 5 in table~1}
\label{fig:Curved2}
\end{figure}

Using type A memristors in series, as in circuit 3, gave an $I$-$V$ profile similar to that for one memristor (compare with figure~\ref{fig:single_mem_new}) other than an unexpected spike near the end. However, putting two memristors in `anti-series' as in circuit 2 gave increased noise and several spiking events, as shown in figure~\ref{fig:Curved2}b. The anti-polarity series interactions causes this richer behaviour. Similarly, the 2-memristor parallel interactions, as shown circuits 4 and 5 show more noise and spiking events, see figure~\ref{fig:Curved2}c and d. Note that only the two memristors in series show the expected spike at the start and when the voltage is switched off (as was seen for a single memristor), therefore only circuit 3 can be considered equivalent to a single memristor.

\subsubsection{Type B (triangular) Memristors in 2-Memristor Circuits}

\begin{figure}[!tbp]
\centering
\subfigure[]{\includegraphics[width=0.49\textwidth]{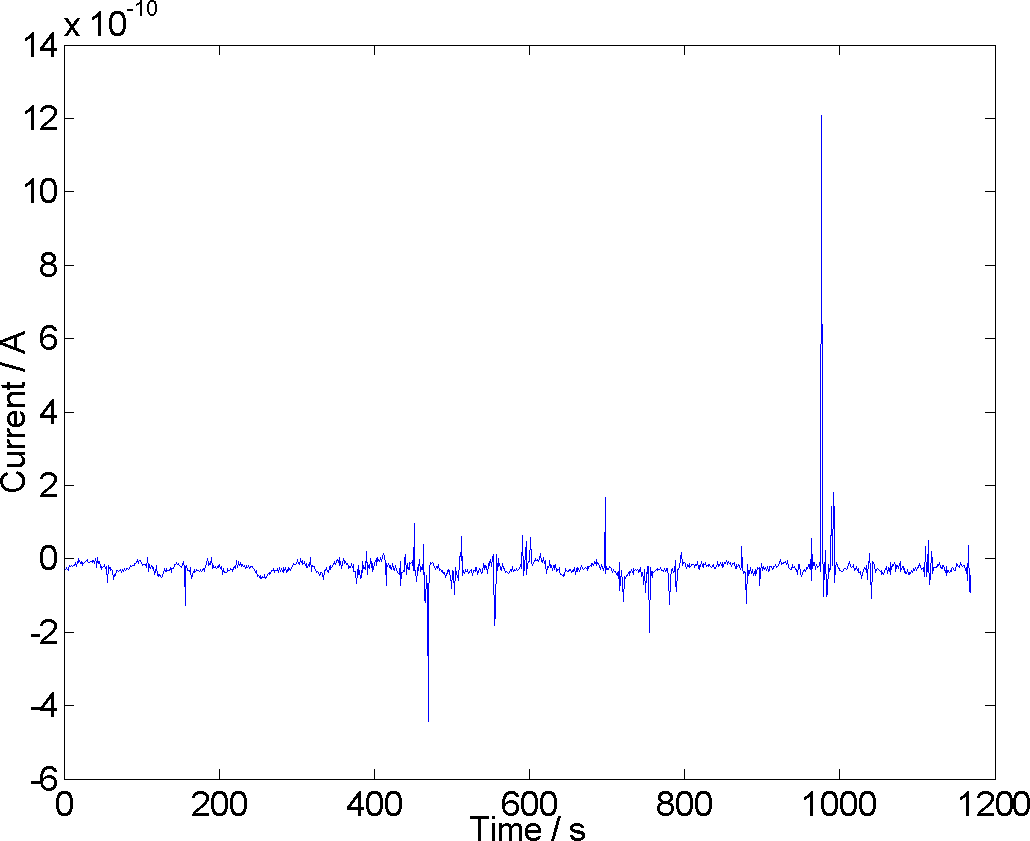}}
\subfigure[]{\includegraphics[width=0.49\textwidth]{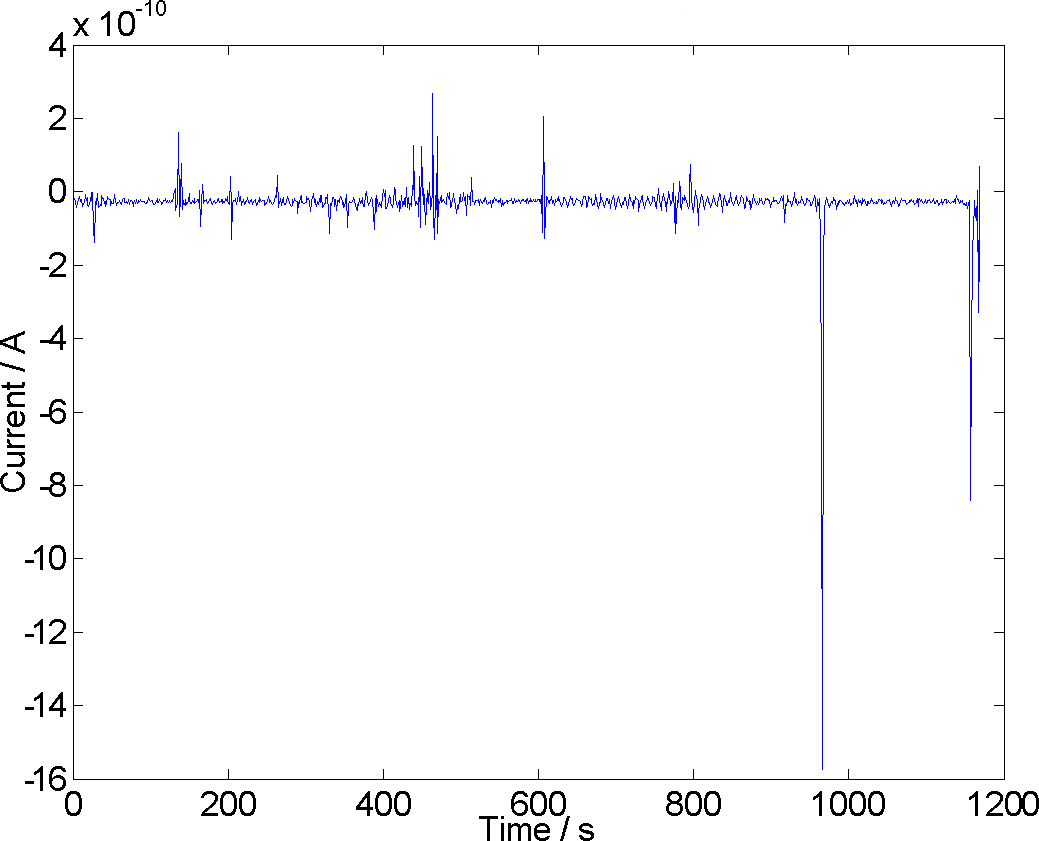}}
\subfigure[]{\includegraphics[width=0.49\textwidth]{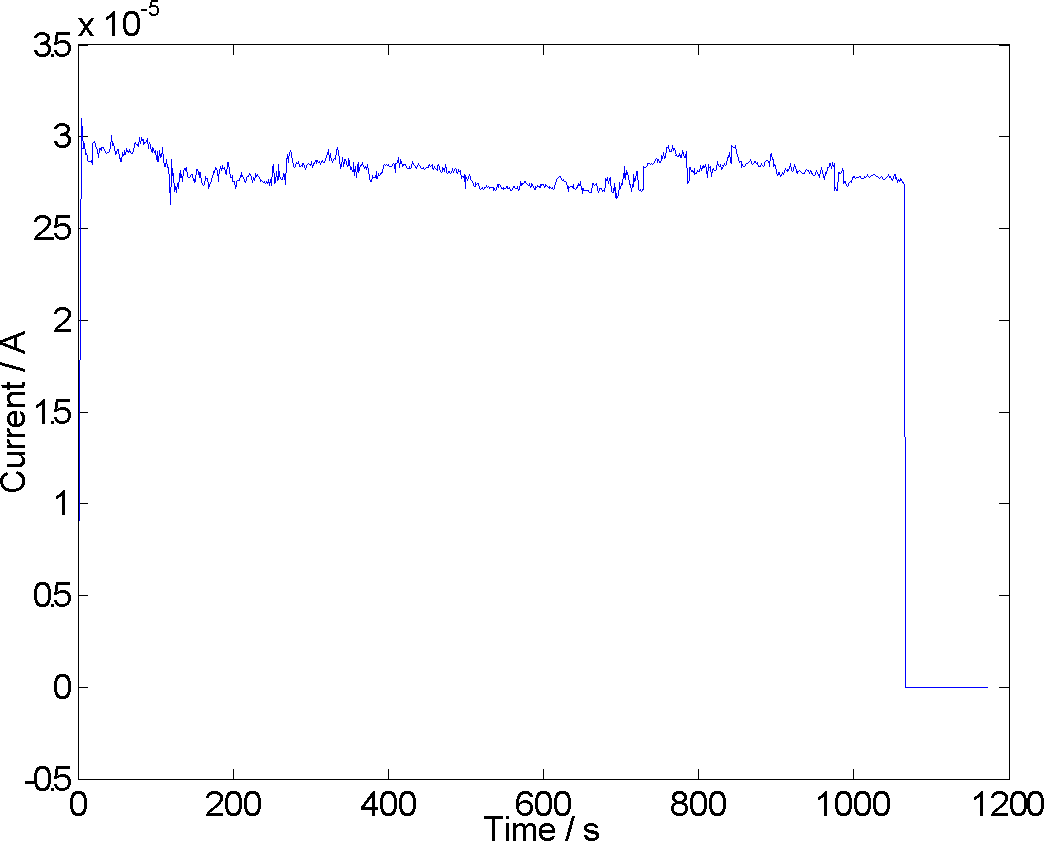}}
\subfigure[]{\includegraphics[width=0.49\textwidth]{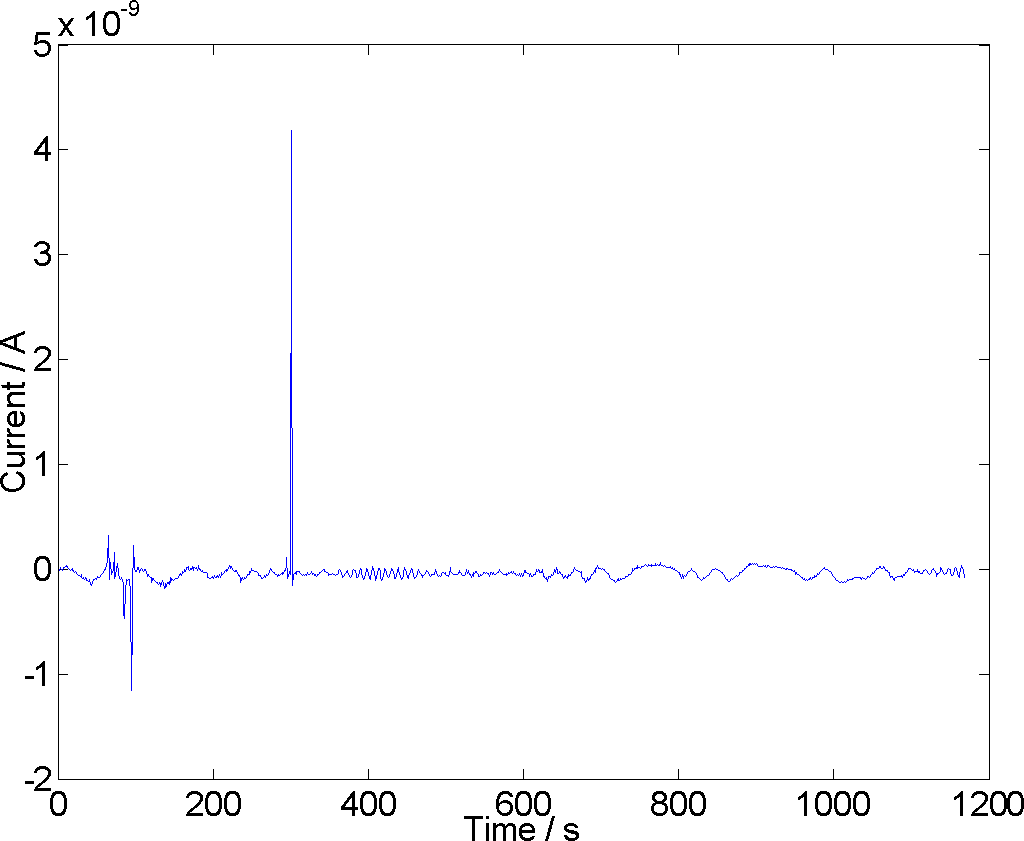}}
\caption{Results for type B (triangular) memristors: a. Two B memristors in anti-series, circuit 2 in table~1, b. Two B memristors in series, circuit 3 in table~1, c. Two B memristors in anti-parallel, circuit 4 in table~1 and d. two B memristors in parallel, circuit 5 in table~1}
\label{fig:Triangular2}
\end{figure}

Figure~\ref{fig:Triangular2} shows the results from the constructions of circuits 2,3,4 and 5 with filamentary memristors. These circuits show a richer behaviour with the emergence of oscillatory type behaviour in circuits 2, 3 and 5. Filamentary memristors in series (circuit 3) do not act like single memristors. Also, in some cases the filaments partially connect, as seen in figure~\ref{fig:Triangular2}c.

\subsection{Three Memristor Circuits}

\subsubsection{Type B (triangular) 3-memristor circuits}

\begin{figure}[htbp!]
 \centering
 \includegraphics[bb=0 0 576 432,scale=0.5,keepaspectratio=true]{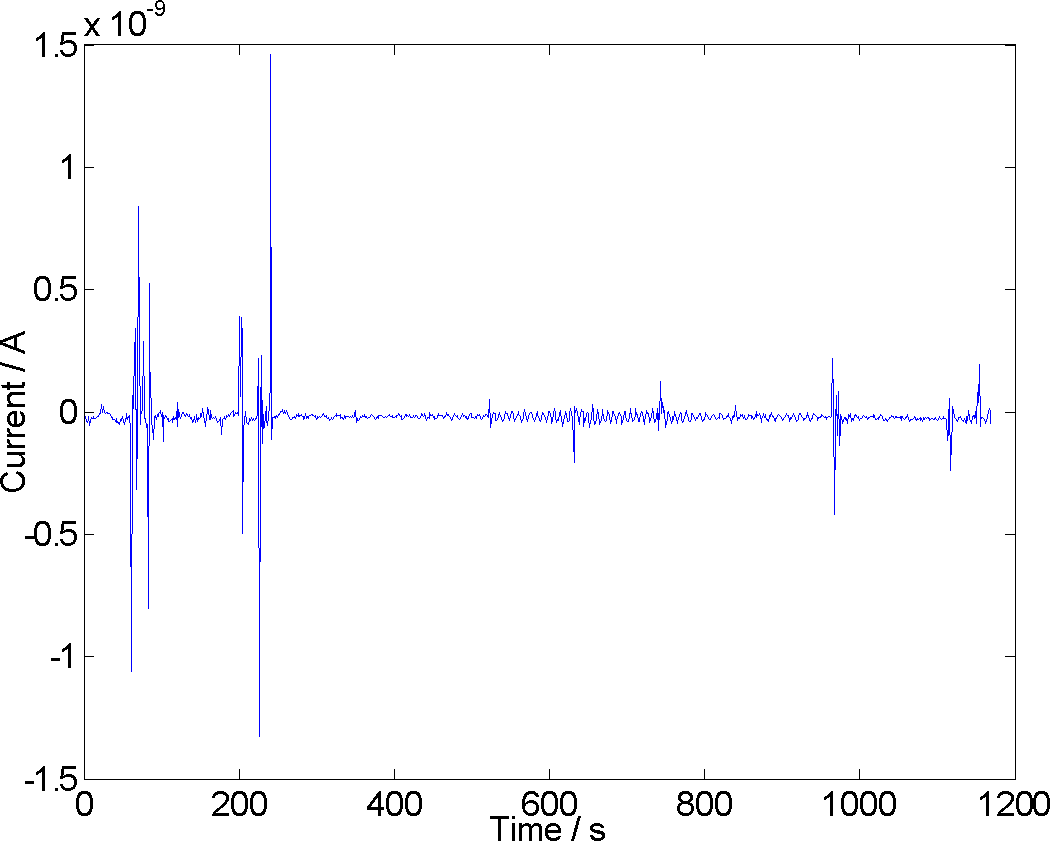}
 \caption{Another typical $I-t$ profile for circuit 6 in table ~\ref{tab:TestCircuits}.}
 \label{fig:MT1p4}
\end{figure}

Typical results for this circuit 7 are given in figure~\ref{fig:101012MT1p1}a. Comparing this with the expected curve in figure~\ref{fig:single_mem_new}a for one memristor shown in figure~\ref{fig:single_mem_new}a, we can see differences. The large spike at the start has vanished, as has the one at the end. We see oscillations in the baseline interrupted by spontaneous spiking. Figure~\ref{fig:MT1p4} shows a later run where we see sections of oscillations of different frequencies. The experiment was repeated to see if there was a repetition in the spiking pattern and thus if the circuits were following long-term periodic dynamics, this was not observed.


\begin{figure}[htbp!]
 \centering
 \includegraphics[bb=0 0 576 432, scale=0.5]{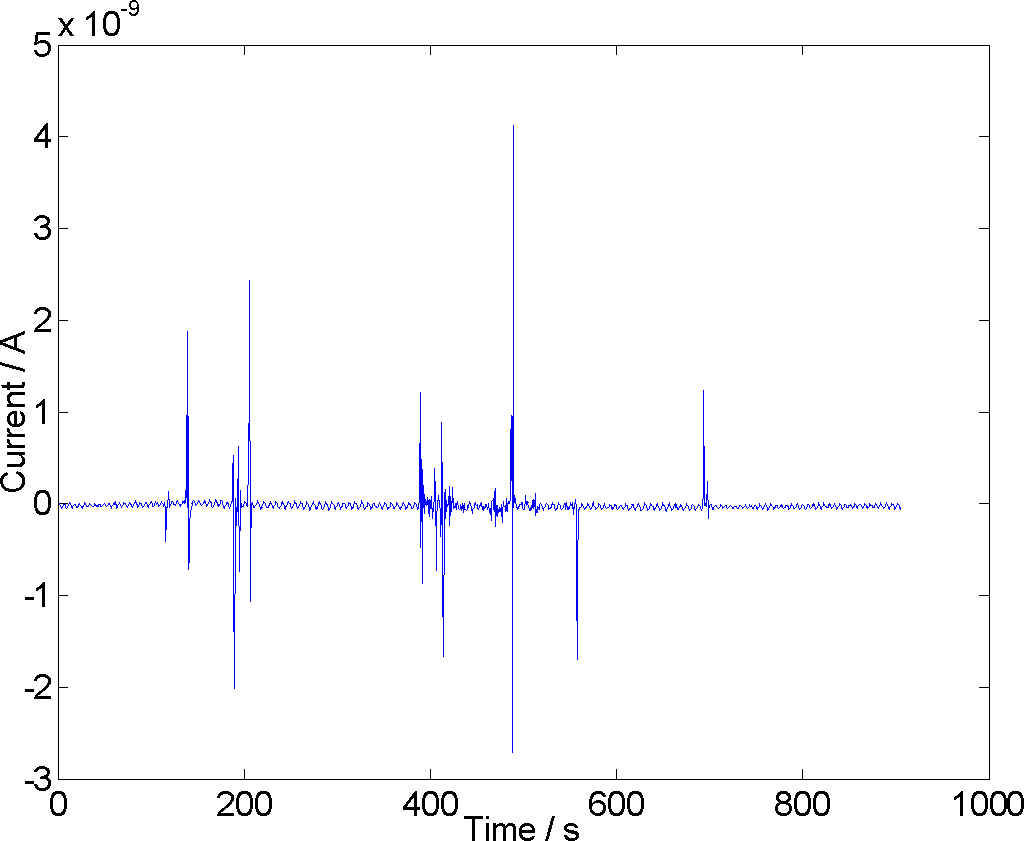}
 \caption{An I-t curve for a very slow sinusoidally varying voltage.}
 \label{fig:MT2}
\end{figure}

We attempted to effect this oscillation by driving it with a very slow a.c. voltage, the results of which are shown in figure~\ref{fig:MT2} (the whole data is for one period). This does not show the expected response for a single memristor (a lagged and distorted sinusoidal current), or any change in the `baseline' as a result of the changing voltage, however the auto-correlation data looks more harmonic than for the constant voltage (figure not included).

\clearpage

\subsection{Type A 3-memristor Circuits}

Do these results mean that multi-memristor circuits do not combine as expected? Not necessarily. We decided to repeat the tests with three type A memristors which are closer to theoretical Chua memristors (type A from paper~\cite{260}). We specifically chose three memristors that had similar looking $I-V$ curves that operated over a similar current range (i.e. were in a similar starting state) to try and decrease the compositional complexity of the circuit.

Figure~\ref{fig:181012MTD7p2t}a shows the same memristors wired up in circuit 7, and the output current looks like a single memristor. For the type A memristors, we found that three memristors in a circuit wired up with the same polarity (i.e. as in circuit 6) behave qualitatively just like one memristor. For circuit 6, we also sometimes see occasional switching events with a decay, see figure~\ref{fig:181012MTD6p1}b which are designated as `w'. The shapes of these individual curves are similar to the overall spike curve seen when the voltage switches, which strongly suggests that the more event-rich behaviour seen in the other memristor systems are to do with interacting switching spikes. 

\begin{figure}[!tbp]
\centering
\subfigure[]{\includegraphics[width=0.49\textwidth]{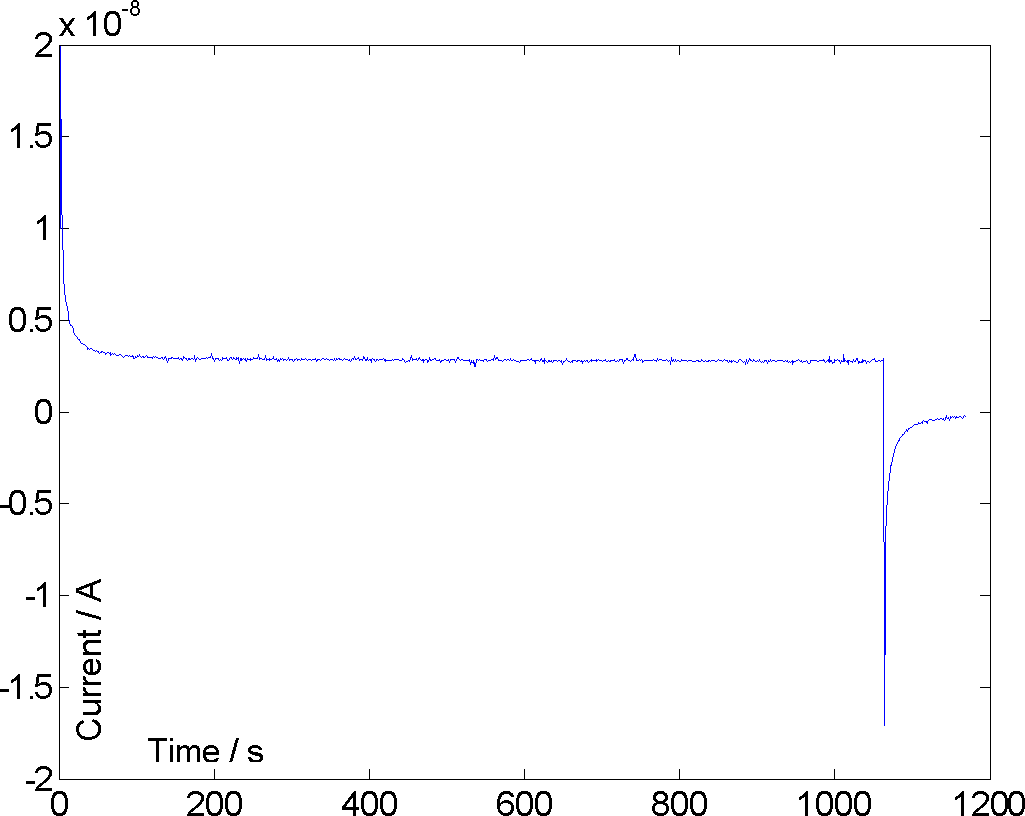}}
\subfigure[]{\includegraphics[width=0.49\textwidth]{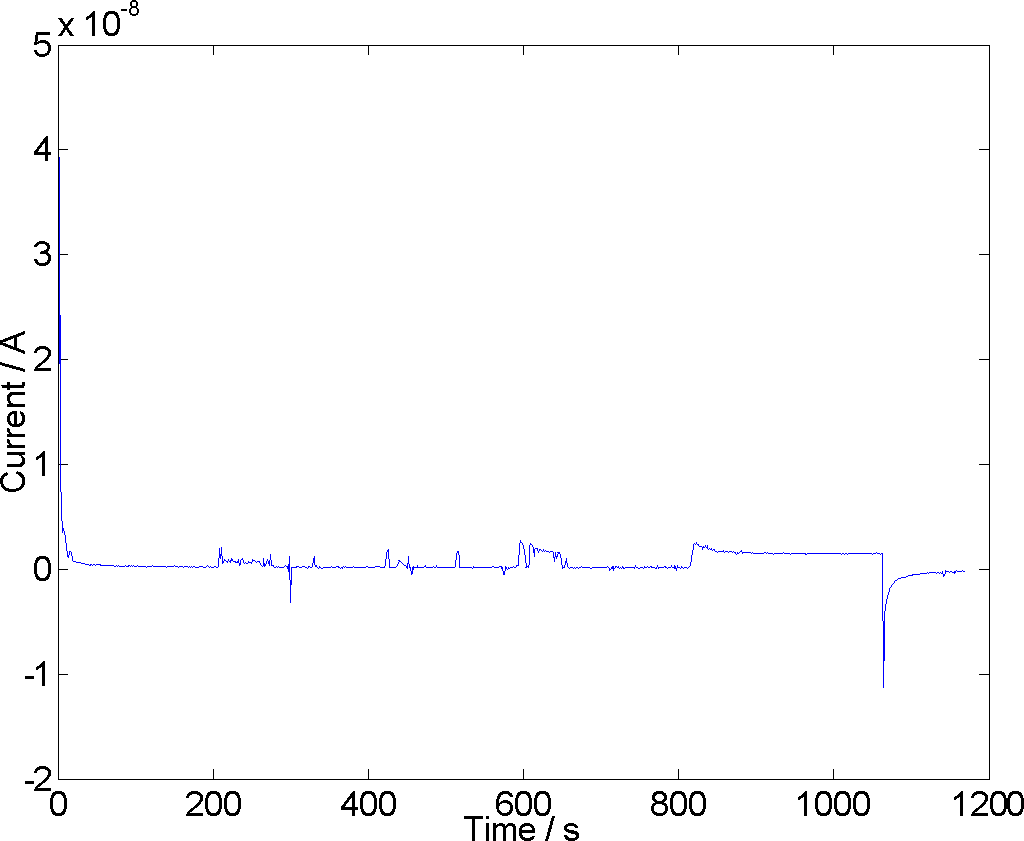}}
\caption{Results for type A memristors: a. three memristors in circuit 7, b. three memristors in circuit 6}
\label{fig:Curved3}
\end{figure}

\section{Statistical Analysis}

\subsection{Ideal Networks}

The periodograms for these data show increasing power with period because the data lacks short time oscillations (other than noise) and has a maximum power at the longer periods because the data is not oscillatory (the long time oscillations are an artifact of the finite experiment length), see figure~\ref{fig:181012MTD7p2t}b. The auto-correlation plots, see for example~\ref{fig:181012MTD7p2t}c, show an exponential drop off far from zero, this is strong auto-correlation indicating that there is a trend present in the data, which is identified as the underlying decay of the current transient. The PACF plots show the minimum persistence of the data (i.e. the order of auto-regression (AR) function required to fit the data), which tails off quickly because this is a simple trend, see figure~\ref{fig:181012MTD7p2t}d which only needs an auto-regression function of order 4 or 5 to encapsulate its dynamics. The presence of spikes in the data causes longer-time persistence in the PACF, the average AR order for `1' 
dynamics was 4.5 compared to 27.6 for `1s' dynamics. Figure~\ref{fig:181012MTD6p1} shows that the presence of  switching effects in ~\ref{fig:181012MTD6p1}a leads to a combined series of exponentials in the ACF (figure~\ref{fig:181012MTD6p1}c) instead of a smooth single exponential, indicating that there are several half-lives in the data due to the switches. 

\subsection{Filaments}

When the filaments in the memristors connected, as in figure~\ref{fig:Triangular2}c, the ACF was a straight line with a very low PACF (of AR(1) for the data in figure~\ref{fig:Triangular2}c) indicating that the data can be modelled as an ohmic filament. This is the case when the memristors connected or almost connected, rendering the circuit essentially a resistor. 

\subsection{Spiking Networks}

The periodograms show a different distribution compared to the ideal networks, they have a large number of low period frequencies  and the power envelope does not increase with increasing period, see for example~\ref{fig:101012MT1p1}b,~\ref{fig:151012MT1p5}b and \ref{fig:241012MTR5p2}b. In some cases the oscillation frequencies can be picked out, but because the system changes over the course of the experiment, the periodograms are not straightforward. The ACF plots help us to understand the dynamics. If there is a periodicity in the ACF (likely one of several interacting frequencies, some examples are figures~\ref{fig:221012MTR6p3}c and \ref{fig:241012MTR5p2}c) then we have sinusoidal waveforms in the data. The decay envelope of these waveforms is a measure of the noise in the system as a proportion of the oscillations and are most clearly visible in the simpler `SpO' results. The ACF patterns are different for these data compared to the single memristor-like data, making the ACF technique a useful 
diagnostic tool. The PACF tends to have higher minimum frequencies for the spiking data, indicating that these data are more complex than the single memristor-like data (as they seemed to be by eye). On average, the spiking networks have a higher AR function order than the ideal networks (29.5 and 30.33 for `Sp' and `SpO'), due to spike correlations introducing a larger persistence in the data. The number of significant ACF lags is smaller for the spiking networks (14 and 5.17 for `Sp' and `SpO' respectively) than for the ideal networks (which are over 50) because there is no background trend in the spiking network data.

\subsection{Is the Spiking Caused by Random Processes?}

For all the data presented in table 2 the Ljung-Box Q-test found that data was non-random, with the exceptions of the experiments in figure~\ref{fig:101012MT1p1} (circuit 7), figure~\ref{fig:241012MTR5p2}  (circuit 5) and circuit 8 (graph not shown), which had respectively 15/50, 18/50 and 25/50 auto-correlation lags might have been the result of a random process. Inspection of the $I$-$t$ curves at these points showed that it was the low power oscillations which were sufficiently noisy that we can not say with 95\% certainty that they are not the result of random processes (and, of course, we expect noise in our experimental system). This does show that although we cannot entirely rule out noisy processes as the cause of our observations we can be reasonably certain that most of the observed oscillations and spikes are not noise. Thus, the high minimum order AR requirement suggests the increased persistence discussed above is due to spike correlation and not a random effect.

\subsection{Circuit (Compositional) `Complexity'}

From the data in table 2 we can draw some conclusions about the circuits. As discussed in the previous section, type A memristors are closer to the Chua memristor and thus act more like an ideal system, in that combinations of them are more likely to behave like a single memristor. All the observed ideal network behaviour came from A type memristors, and accounted for 6/10 A type circuits. Similarly, all the `fi' dynamics (those which are thought to be the result of filament connecting) arose in circuits made with B type memristors (accounting for 3/12 B type circuits) which is expected as these are filamentary memristors. 

The number of significant ACF lags increases in both A and B type circuits with an increasing number of memristors. For 2 memristors, A circuits have an mean of 18.5, B circuits 14.75, overall 16.62. For 3 memristors, A circuits have a mean of over 50, B circuits have 22.875, which is because 2 memristors are overwhelmingly likely to be `1s' type dynamics. Concentrating on just the spiking networks, we have an increase in mean significant number of lags from 3 for the 2-memristor networks to 14 for the 3 memristor networks, suggesting that 3 memristor circuits have a significantly higher persistence. The order of AR function is 22.8 for 2-memristor circuits and 33.4 for 3-memristor circuits (due to the fact that `Sp' dynamics are more complex and observed more often in 3-memristor circuits), quantifying the amount that 3-memristor circuit dynamics are more complex (which accords with what was observed). This demonstrates that the ACF and PACF data does seem to identify and quantify the concept of `complexity'
 in the circuit dynamics. 

We can perhaps take this complexity analysis further. Table 1 includes a count of the anti-series and anti-parallel interactions as an attempt to roughly quantify the compositional `complexity' of the circuit (where we are taking complexity to mean the likelihood of rich dynamics and not a measure of the time taken to solve a problem). For the 12 spiking data sets, the mean PACF for the circuits can be used to order them based on the complexity of the dynamics. The three circuits with 0 interactions (i.e. all the memristors are wired up with the same polarity), numbers 3 and 5 have mean minimum AR functions of order 28.5 and 21.5 respectively. Circuit 6 with 3 interacting memristors of the same polarity do not generate spikes, perhaps due to the 3 memristor networks stabilising any arising fluctuations the way that anti-polarity networks propagate them. The circuit with 1 anti-series interaction requires an AR of order 14, and the circuit with 1 anti-parallel interaction requires an AR of order 45, agreeing 
with our expectation that anti-parallel circuits would give richer behaviour than anti-series. In the three-memristor circuits, as we suspected, having 2 different types of anti-polarity interaction (i.e. anti-series and anti-parallel) gives richer dynamics requiring an AR of order 40.33 to fit, whereas a circuit with 2 anti-parallel (circuit 8) required 29.6 (however, this was a smaller set so this result has a larger standard error). 

It is interesting to compare this work to our ongoing work with neural cell cultures connected to single memristors~\cite{Ne0}. In those experiments, the single memristor underwent `1', `w' and `1s' type behaviour, demonstrating that this behaviour can arise from single memristor circuits. These experiments also showed noisy spiking that is believed to be from the spiking of the neural cells due to its different character (oscillations and bursting spikes were not observed in the joint system). A more-thorough analysis of those $I$-$t$ curves using the techniques presented here will be forthcoming.





\begin{table}
\begin{tabular}{|c|c|c|c|c|c|c|}
\hline
Experiment 	& Circuit 	& No. 		& Type 		& No. of 	& PACF 		& classi-\\
No.	 	& No. 		& of 		& of 		& significant	& minimum	& fication\\
		& 		& memristors	& memristors	& ACF lags	& AR order	&  \\
\hline
-- 		& 1		& 1		& A		& 43		& 5  		& 1\\
\hline
\hline
1 		& 2		& 2		& A		& $>$50		& 42 		& 1s\\
2 		&2		& 2		& B		& 5		& 14 		& SpO\\
\hline
3 		& 3		& 2		& A		& 3		& 46 		& SpO\\
4 		& 3		& 2		& B		& 2		& 11 		& Sp\\
\hline 
5 		& 4		& 2		& A		& 16		& 45 		& SpO\\
6 		& 4		& 2		& B		& $>$50		& 2  		& fi\\
\hline
7 		& 5		& 2		& A		& 4		& 36 		& SpO\\
8 		& 5		& 2		& B		& 1		& 7	  	& SpO\\
\hline
\hline
9 		& 6		& 3		& A		& $>$50		& 37 		& 1s \\
10 		& 6		& 3		& A		& $>$50		& 5  		& w \\
11 		& 6		& 3		& A		& $>$50		& 4  		& 1s \\
12 		& 6		& 3		& B		& $>$50		& 1		& fi\\
13 		& 6		& 3		& B		& $>$50		& 50 		& fi\\
\hline
14 		& 7		& 3		& A		& 50		& 3		& w\\
15 		& 7		& 3		& A		& $>$50		& 5		& 1\\
16 		& 7		& 3		& A		& $>$50		& 4		& 1\\
17 		& 7		& 3		& B		& 6		& 31		& Sp\\
18 		& 7		& 3		& B		& 18		& 47		& Sp\\
19 		& 7		& 3		& B		& 13		& 33		& Sp\\
\hline
20 		& 8		& 3		& B		& 12		& 47 		& Sp\\
21 		& 8		& 3		& B		& 2		& 34 		& SpO\\
22 		& 8		& 3		& B		& 33		& 8 		& Sp\\
\hline
\end{tabular}
\caption{Results from repeats. Key :- `Sp': bursting spikes and oscillations; `SpO': Oscillations with single spikes; `fi': filament connecting; `1': single memristor-like; `1s' single memristor like with single spikes; `w': switching single memristor.} 
\label{tab:2}
\end{table}

\section{Conclusion}

We have introduced the concept of compositional complexity as a way of comparing circuits' compositions -- to our knowledge there is not an existing framework to do this (although several rules of thumb are known). This work shows that the intuitive idea about which circuits are more complex does seem to work for predicting increased complexity in the circuit dynamics, and that partial autocorrelation function and autocorrelation function analyses are a useful tool in quantifying this complexity. Specifically, compositional complexity is a measure of the number of memristors and the interactions between them; for 2 memristors the most complex configuration is anti-parallel and in series the least complex.

The idea of compositional complexity was invented to find a metric to compare circuits with the same function -- beyond counting the number of components used or measuring the area of silicon covered, which are not fair comparisons for memristors because they can do more than other components~\cite{UCNC,Spc2} and one of their advantages is that they can be made much smaller. We would like to separate the gains caused by clever circuit design from those caused by technological progress shrinking of circuit size. Putting this idea on a more quantitative standing is an area of current research.

We have shown that anti-parallel interactions give rise to more complex behaviour, specifically an increased likelihood of spiking and more complex spiking dynamics. This result demonstrates that the chaotic dynamics seen in simulations~\cite{272} is likely due to the sub-circuit of two anti-parallel memristors rather than numerical effects. Our results could also suggest that the repeating spiking oscillations seen in the neuristor~\cite{252} may arise from the anti-parallel sub-circuit rather than the interaction between a memristor and a capacitor. We know from the data presented here that the spikes are not random, but further work will be required to demonstrate if they are chaotic in nature -- the fact that 3-memristor networks possess a richer, more persistent and quantitatively different dynamics to 2-memristor networks is suggestive (because chaotic systems need at least 3 state variables).

We have classified our results into two types: The first are ideal networks which show voltage-switch-related spikes followed by smooth exponential decay. The second are switching networks, which lack the voltage-switch-related spikes, have a flat baseline and possess both oscillatory dynamics and bursting spikes. The ideal networks combine like resistors, in that a network of only memristors addressed by their joint 1-port entry is indistinguishable from a single memristor. The spiking networks add up differently and emergent dynamics would make it easy to tell that there was more than one memristor present if we were given a `black-box' circuit and asked to identify whether it contained a single memristor or a network (this is the sort of thought experiment the phrase `addressed by their joint 1-port entry' describes~\cite{Chua1969}).

This work demonstrates some methods for controlling which behaviour is observed in a memristor network. To build an ideal network we require A-type or ideal memristors (i.e. described by the base memory-conservation theory~\cite{F0c} and Chua's constitutive relation~\cite{14}), a low degree of compositional complexity and a smaller network. To build a spiking network we can use either ideal or filamentary memristors (those which require the extension of a ohmic filament~\cite{254}, equivalently those which are memristive systems~\cite{84} with the state of filament as a second state variable), a network above 2 memristors (for the most stable dynamics) and a high degree of compositional complexity.

What is the underlying cause of the spikes? These spikes are widely observed in memristors and referred to as current transients -- which is a description of the dynamics rather than a cause -- the cause is often ascribed to capacitance in the circuit. These networks were tested without a capacitor in the system, but there is of course parasitic capacitance in the circuit and all real devices include some measure of capacitance, resistance and inductance. Furthermore our devices have aluminium electrodes which are known to act as sources and sinks of oxygen ions~\cite{178,175,177,189}, which can stabilise switching~\cite{229,159}, and can be thought of as a slow, ionic capacitor. 

However, the spikes can be fitted by memristor theories (both ours~\cite{F0c} and others~\cite{15}, see~\cite{SpcJ}), which suggests that they may arise from the memristance in the devices. When it first appeared, the memristor was thought of as an a.c. device and the question of the d.c. response was not addressed. But obviously, the ions within the memristor will drift to voltage regardless of whether it is varying or constant (with the proviso that if the voltage varies too fast the ions won't have time to drift one way significantly before changing direction; this is covered by the part of the memristor definition which described the hysteresis as the frequency tends to infinity, see~\cite{14,279}). Thus, we expect there to be a time-varying response to a d.c. current, and the observed current spike is a compelling candidate. We have experimentally demonstrated that this spike shape is related to frequency-related hysteresis changes~\cite{hystc}, a fingerprint of the memristor. This fact is suggests that 
the spikes are memristance, but the question is not conclusively answered yet.  

We shall now take the position that the spikes are mainly due to memristance rather than capacitance. A question of interest is then why does the current equilibrate (which we know it does, see~\cite{SpcJ}). This is something we are investigating and early results are presented in~\cite{Spc2}. Due to this behaviour, we have called the spikes the short-term memory of the memristor and have shown that single memristor-like spikes can interact within this time window and can be used to implement logic gates and perform addition~\cite{UCNC,Spc2}. If the memristors in the circuit possess a short-term memory it could explain the cause of the emergent behaviour. As these tests were done across the whole circuit, we do not know how the state of the 3 memristors vary individually. If there is a time delay in a voltage induced current spike propagating across the network, then the response of a memristor to it would change the voltage across another memristor by $\Delta V$. If these changes are not synchronised then 
these small $\Delta V$'s can move around the network causing the individual memristors to spike and propagate a different $\Delta V$ (remember the entire network is subject to a constant voltage so any small change in resistance on one memristor will affect the voltage across the others): this situation is called the `roving $\Delta$V in~\cite{Mu0}. This idea explains the loss of the voltage spikes seen in ideal networks, the dampening of the slow a.c. voltage in figure~\ref{fig:MT2} and the sudden emergence of bursting spikes from an almost flat baseline that has been observed (see the control in~\cite{Ne0}). Thus, the oscillations can then be explained as the `ringing' of the network that results from constructive and destructive interactions of spikes as a $\Delta V$ is passed around and the bursting spikes would occur when the spikes constructively interact. If the bursting spikes were caused by this sort of system, it would explain the correlation between spikes that has been observed. Finally, the $\
Delta V$ would be more likely to become unsynchronised if the network contained more sources of delay and difference, such as those which would be caused by introducing more memristors, more  types of memristors, mismatched polarity, increased numbers of junctions, and so on. These sources of delay match the conditions under which spiking networks were observed as reported in this paper. This is an explanation, however it is not the only one we have entertained. Another explanation is to think of the boundary between on and off resistance material as continually oscillating slightly around equilibrium as ions diffuse around: it could be the interaction of these individual oscillators that causes the oscillations in the network (as is seen in other systems our group studies such as \textit{Physarum}~\cite{AndysBook} and BZ reactions~\cite{336,337}) and the interaction of oscillations could cause the bursting spikes. 

Are these spiking networks brainlike? In this paper we have shown unexpected, emergent properties arise from simple memristor-only networks, which has not been observed before. The pattern of bursting spikes seen is qualitatively similar to those observed in neural bursting spikes. Brainwaves are oscillatory behaviour that are observed and known to be correlated to neural action, but their exact cause is not known. They are thought to be related to the neural voltage spikes (which can encode and process information) and perhaps arise from interaction of these spikes in the network, but there are unanswered questions: which spike patterns give rise to which frequencies; which aspects of the neural network causes which aspects of the dynamics and whether they are an observable side-effect of information processing that arises from the material or actually transmitting useful information. Memristors are similar to braincells because they also involve ionic conduction, have a short-term memory and possess 
spiking dynamics; they differ in that (our) memristors give current spikes in response to voltage whereas the brain operates on (ionic) current-controlled voltage spikes and memristors are simpler in operation than the components of the brain. Thus, memristor networks can act as a simplified hardware model for the brain and we expect that constructing memristor networks can help us understand where electro-physiological phenomena such as brainwaves come from and what they might mean.

The future directions for this work have been described above, but what are the future uses for this technology? If memristor spiking networks are operating under similar principles to those in the brain, it is possible that spiking memristor circuitry would be a good choice for brain-machine interfaces for use in neural prosthetics, disease treatment (e.g. a memristor-based circuit breaker for epileptic fits or a spike generator for dealing with the symptoms of Parkinson's disease) and functionalised prosthetics (i.e. artificial limbs that interface directly with the nervous system). We have started to investigate this by combining spiking memristor networks with spiking neural cell culture to see if the two spiking networks can influence each other electrically~\cite{Ne0}. Another area of interest is biomimetic robotics where spiking memristor networks could be used to process sensory inputs and control a robot. Finally, a spiking memristor computer may be capable of a more bio-inspired approach to 
computing and could prove better at task at which biology excels, such as pattern recognition, fuzzy processing, learning and so on.

\begin{figure}[!tbp]
    \centering
    \subfigure[]{\includegraphics[width=0.49\textwidth]{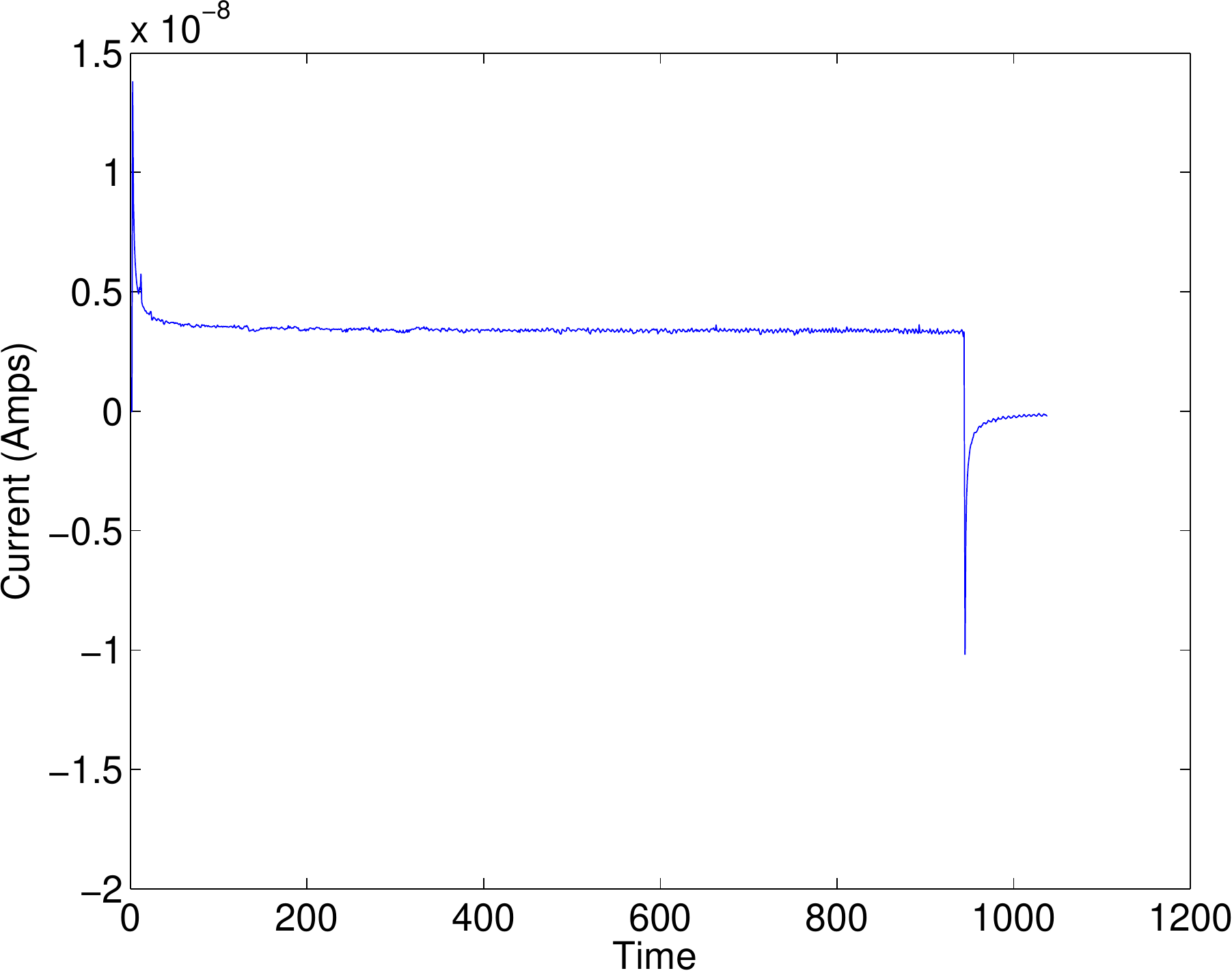}}
    \subfigure[]{\includegraphics[width=0.49\textwidth]{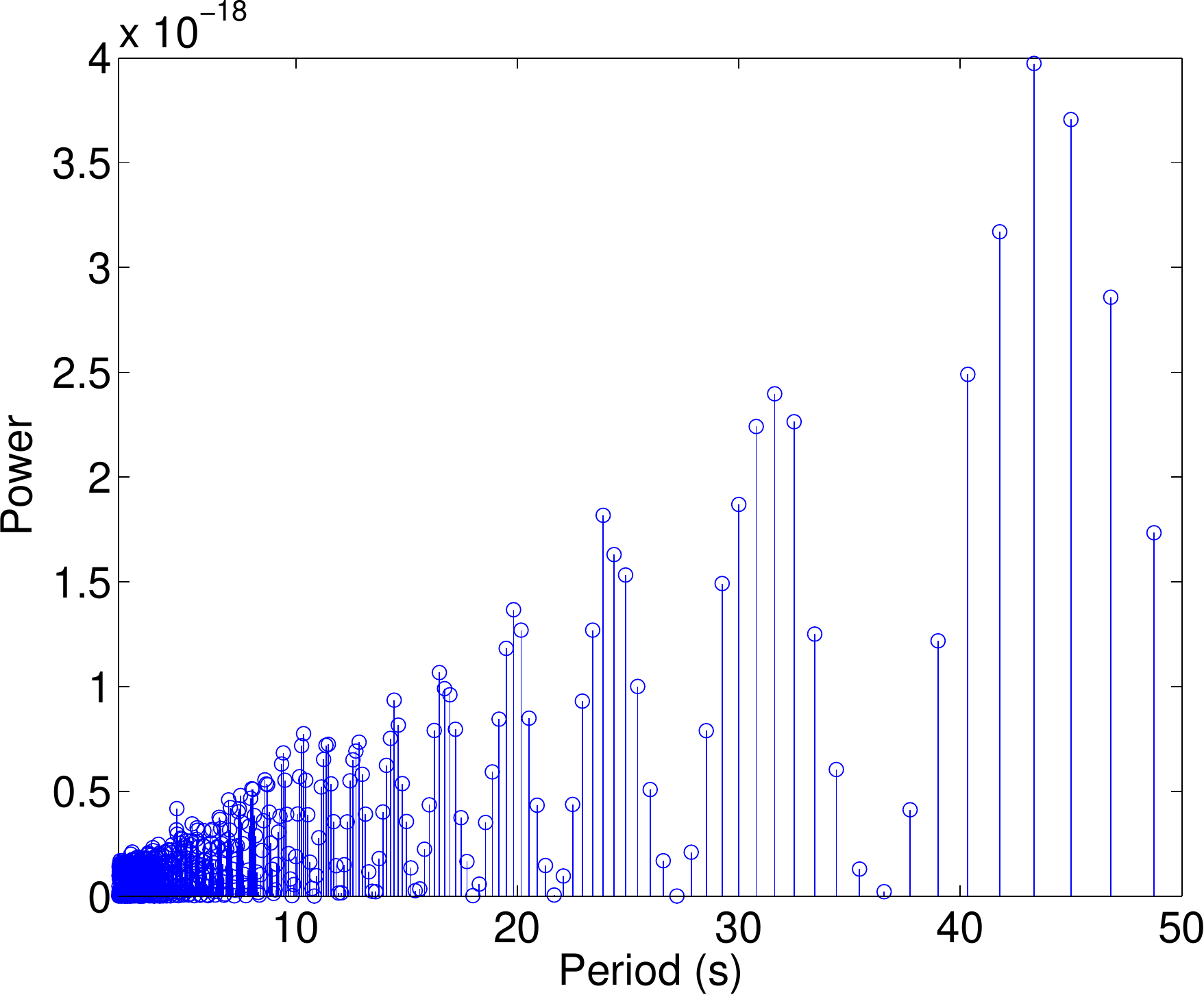}}
    \subfigure[]{\includegraphics[width=0.49\textwidth]{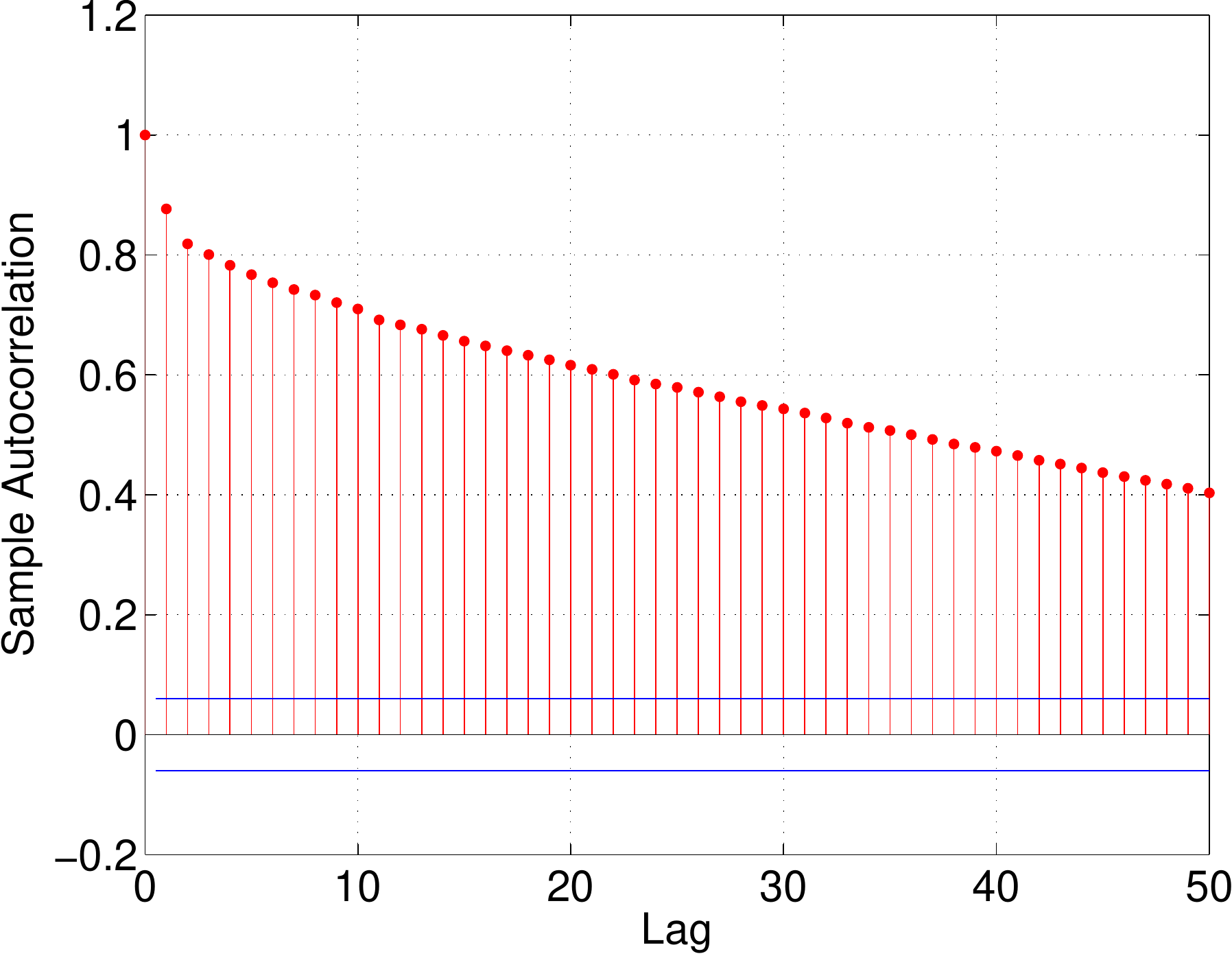}}
    \subfigure[]{\includegraphics[width=0.49\textwidth]{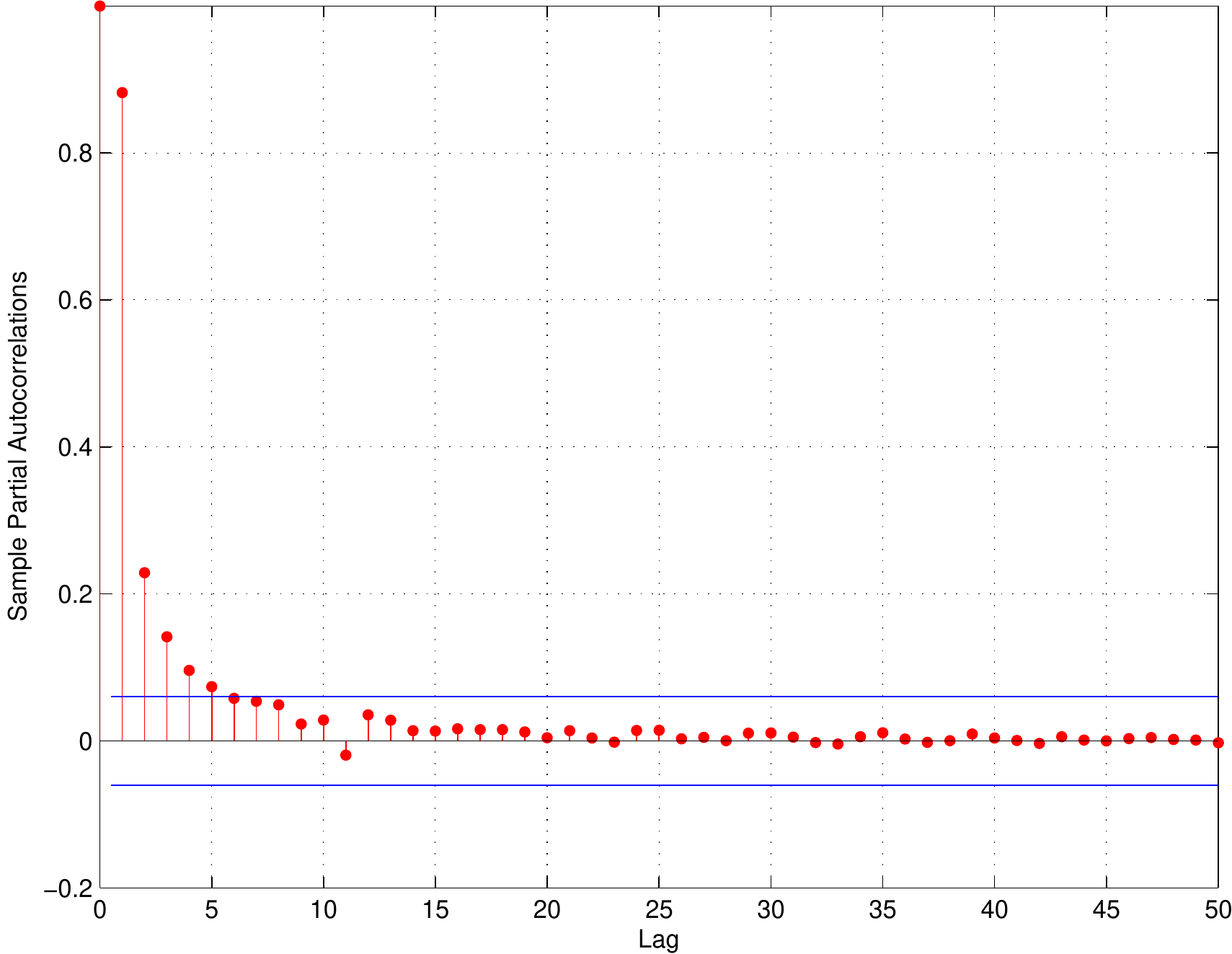}}
    \caption{Single memristor data: a: $I-t$ curve in response to constant voltage; b: a periodogram; c: autocorrelation function; d: partial autocorrelation function. The ACF shows an exponential trend, the PACF shows this is a simple system.}
    \label{fig:single_mem_new}
\end{figure}

\begin{figure}[!tbp]
    \centering
    \subfigure[]{\includegraphics[width=0.49\textwidth]{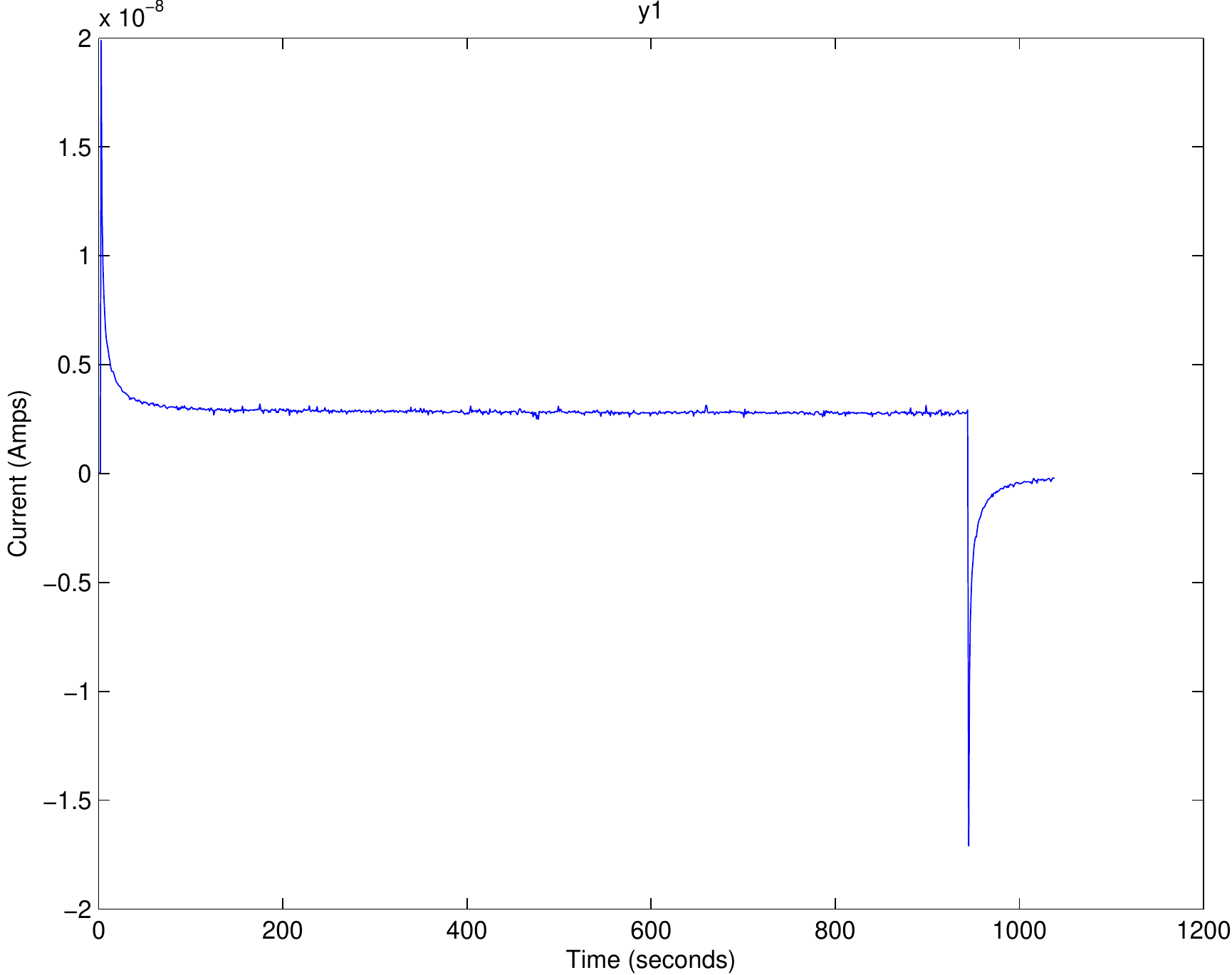}}
    \subfigure[]{\includegraphics[width=0.49\textwidth]{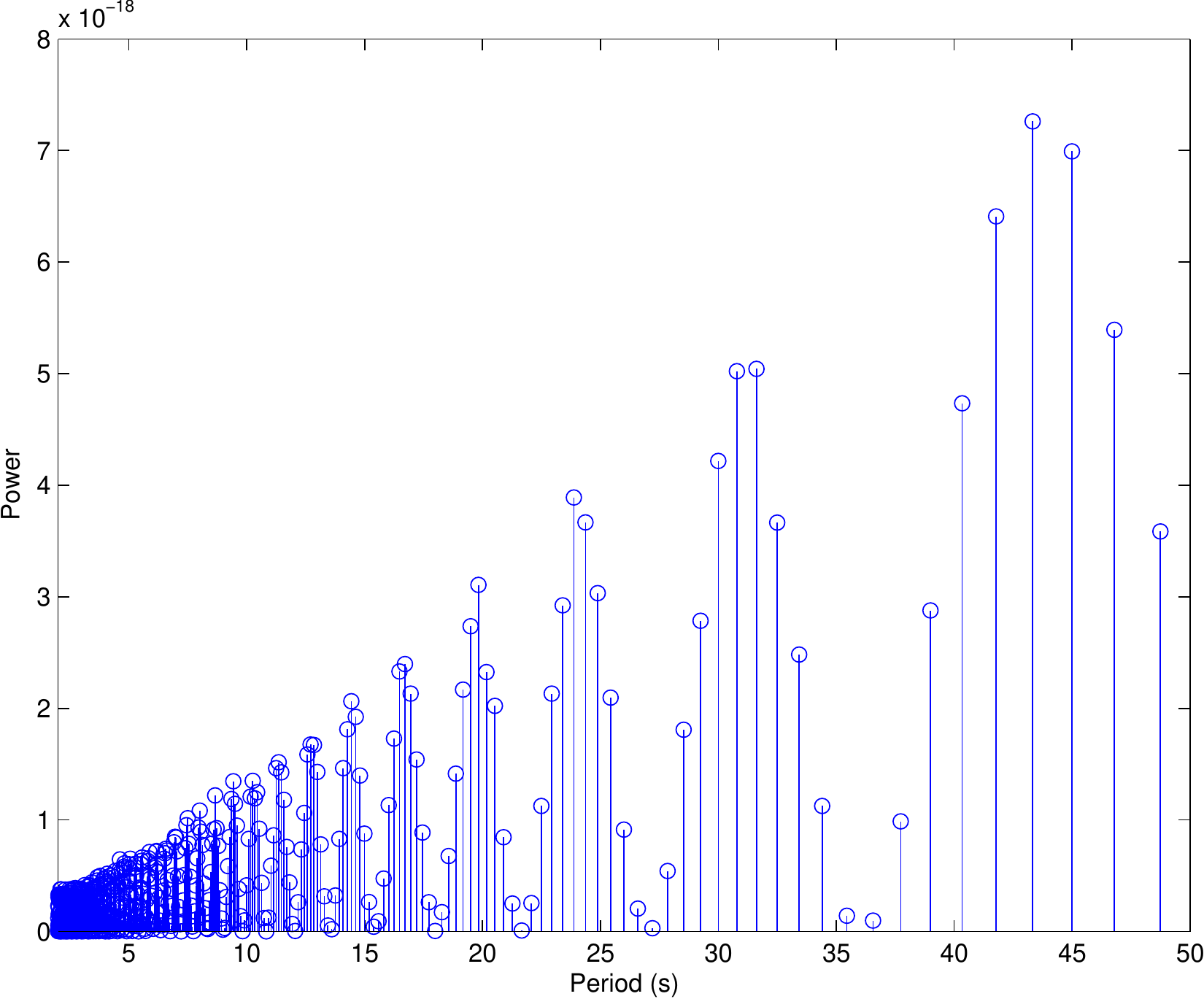}}
    \subfigure[]{\includegraphics[width=0.49\textwidth]{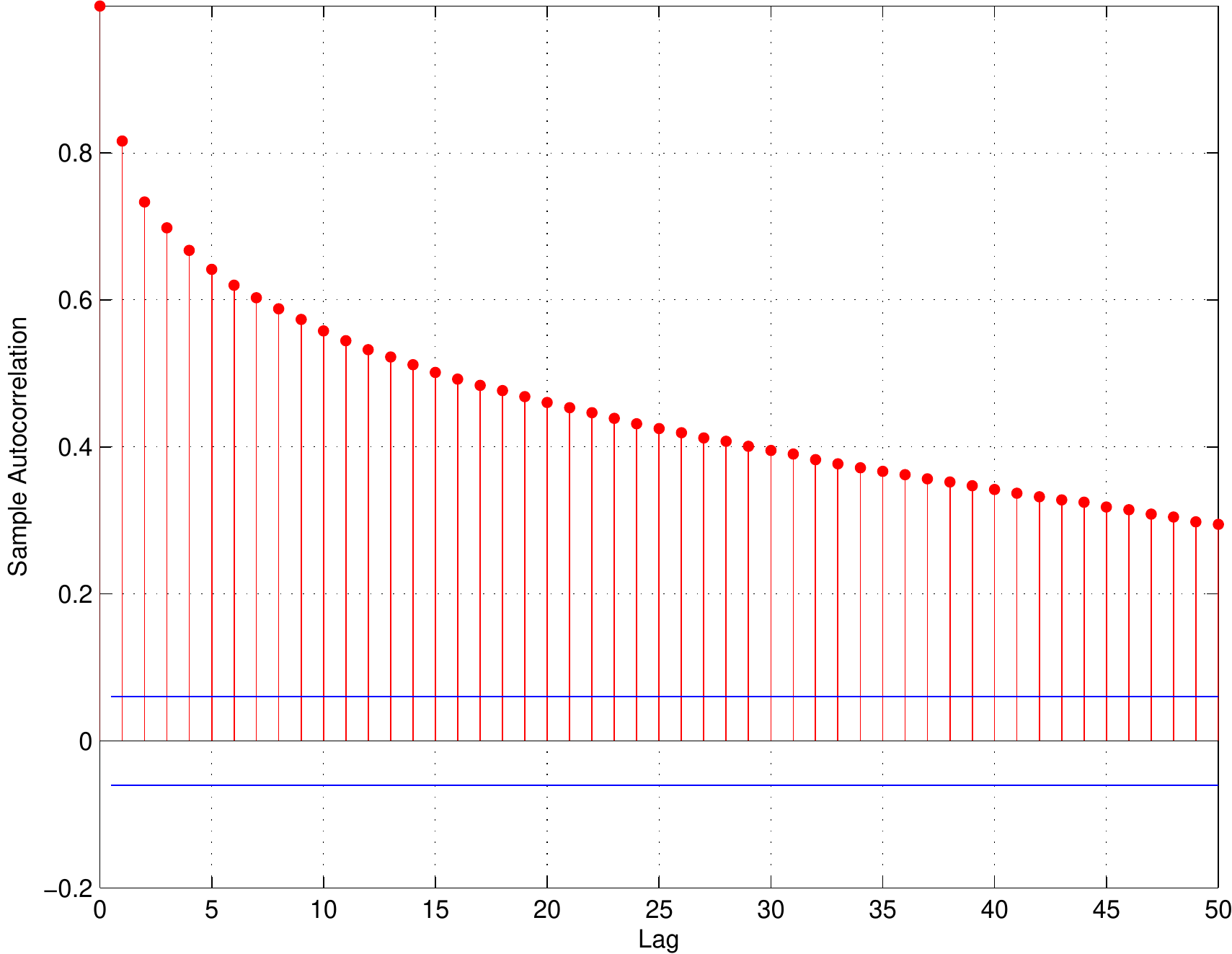}}
    \subfigure[]{\includegraphics[width=0.49\textwidth]{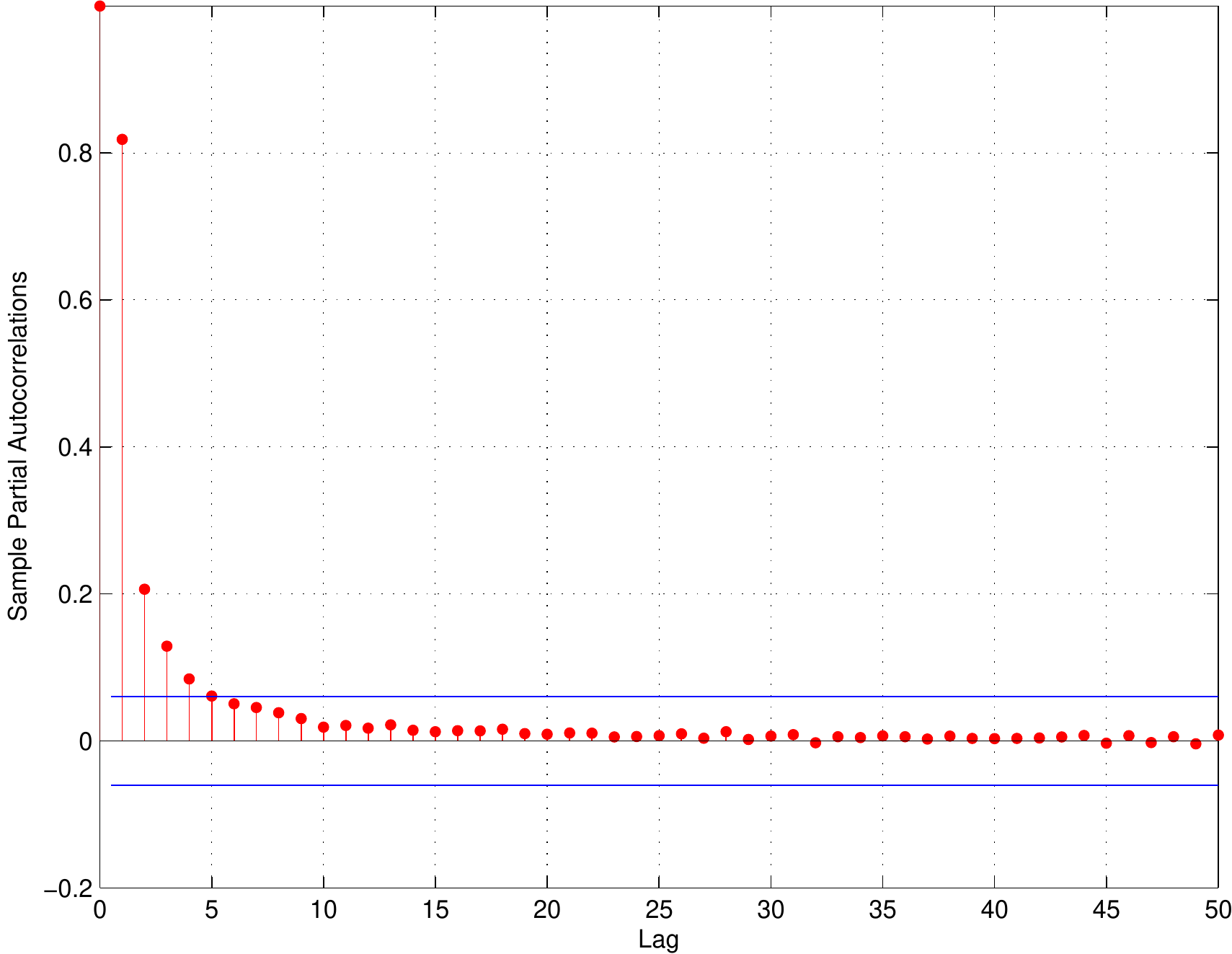}}
    \caption{Ideal network example of 3 A-type memristors in circuit 6: a: $I-t$ curve in response to constant voltage; b: a periodogram; c: autocorrelation function; d: partial autocorrelation function. This demonstrates that using similar memristors with low circuit complexity gives the expected result.}
    \label{fig:181012MTD7p2t}
\end{figure}

\begin{figure}[!tbp]
    \centering
    \subfigure[]{\includegraphics[width=0.49\textwidth]{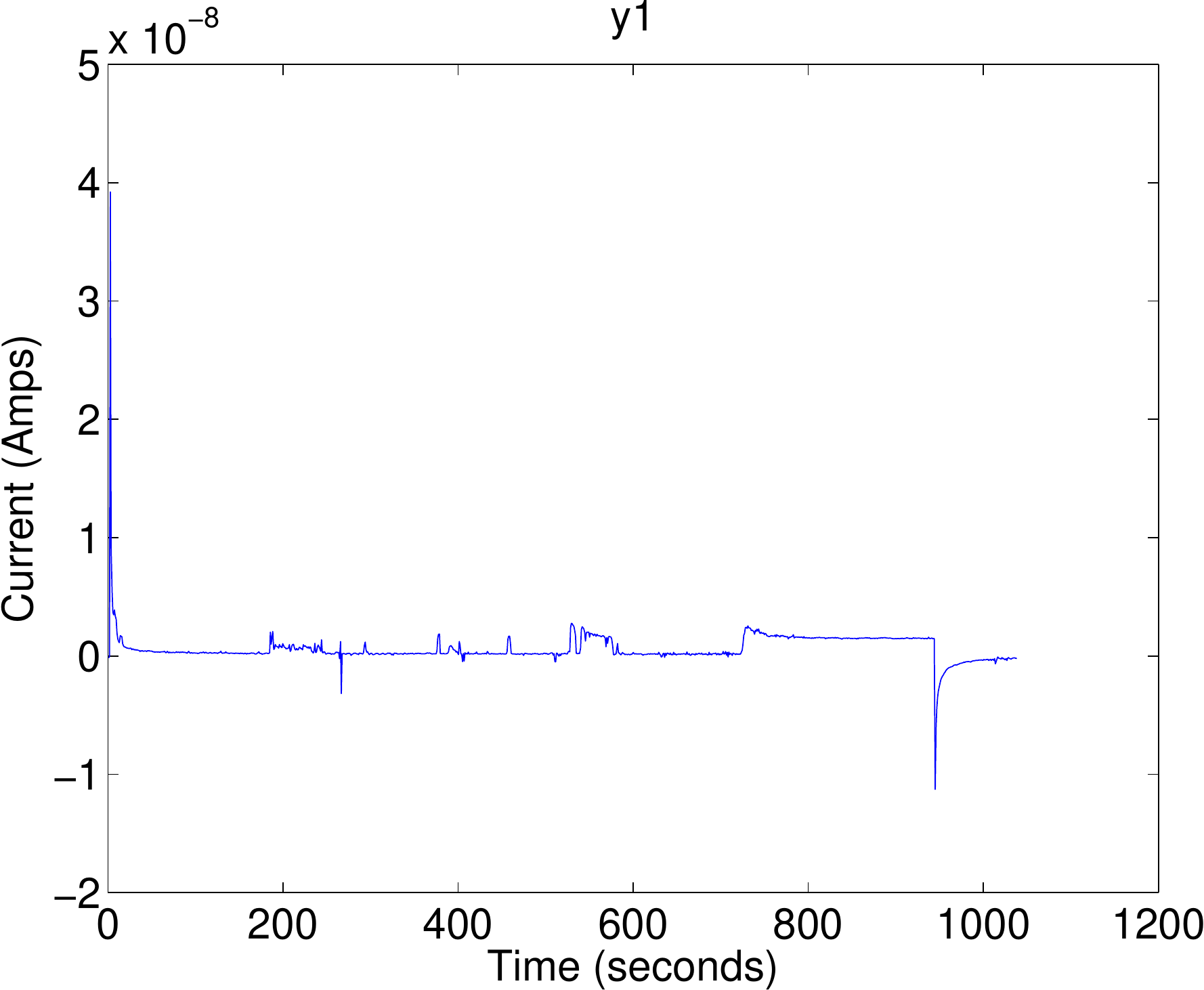}}
    \subfigure[]{\includegraphics[width=0.49\textwidth]{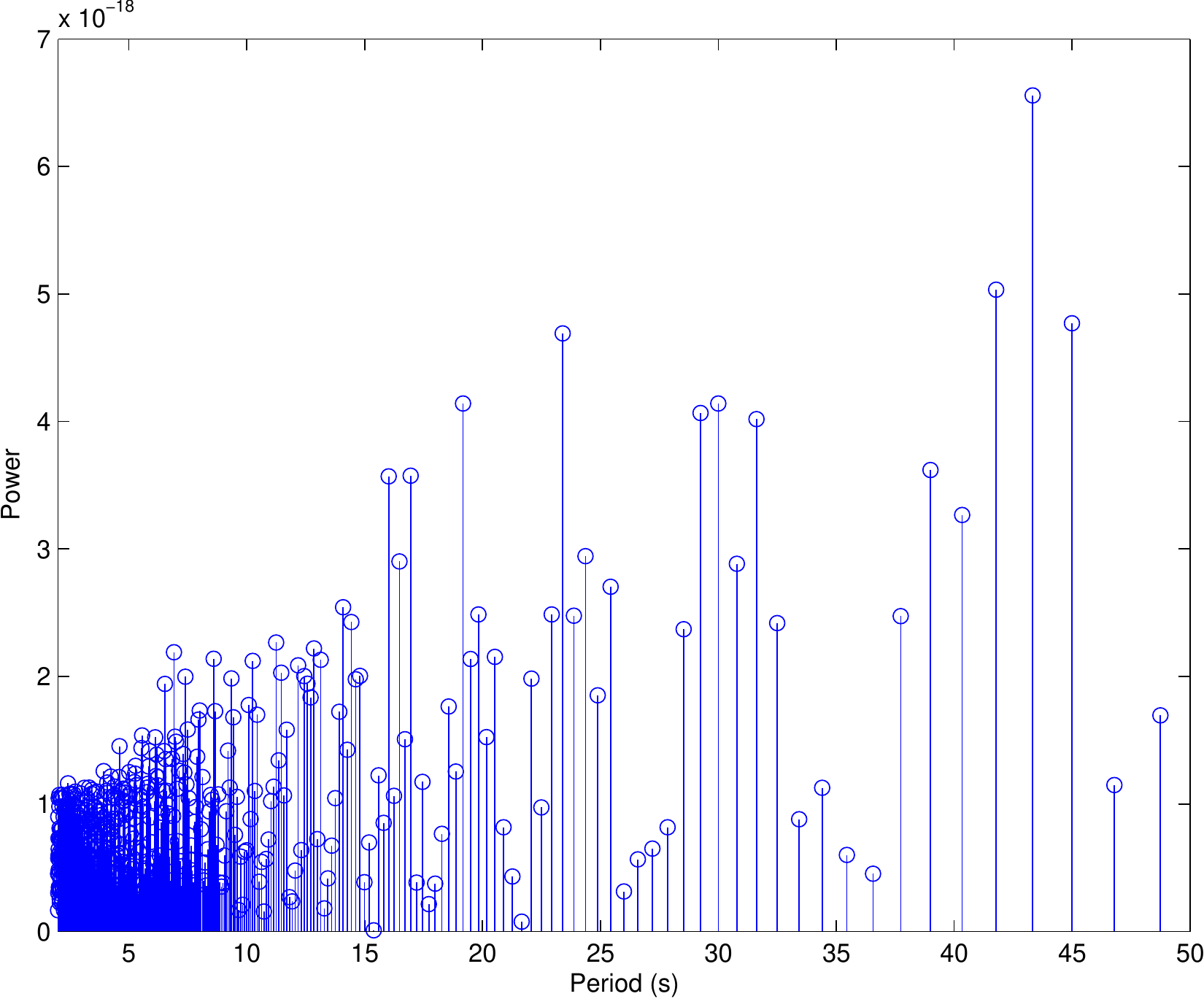}}
    \subfigure[]{\includegraphics[width=0.49\textwidth]{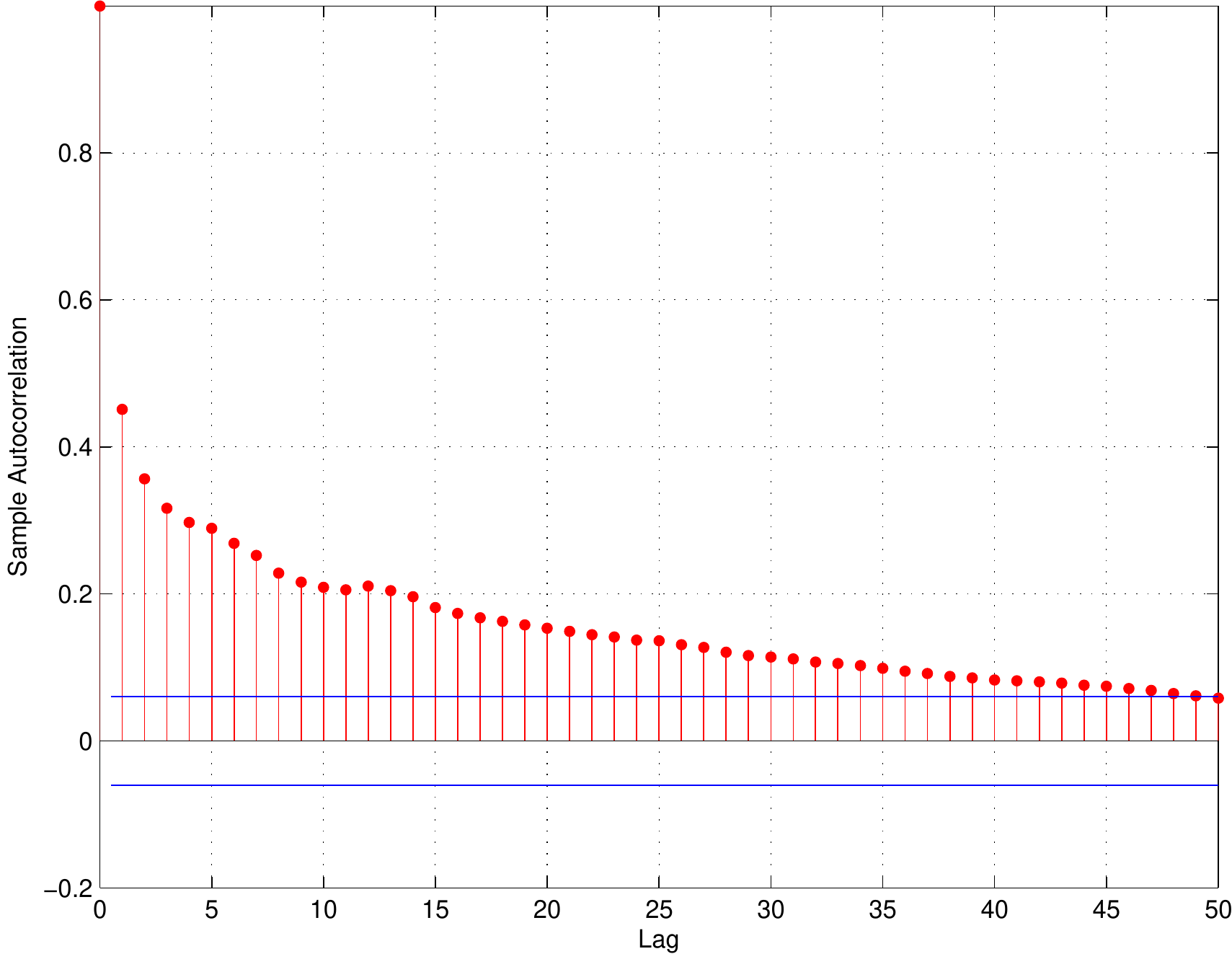}}
    \subfigure[]{\includegraphics[width=0.49\textwidth]{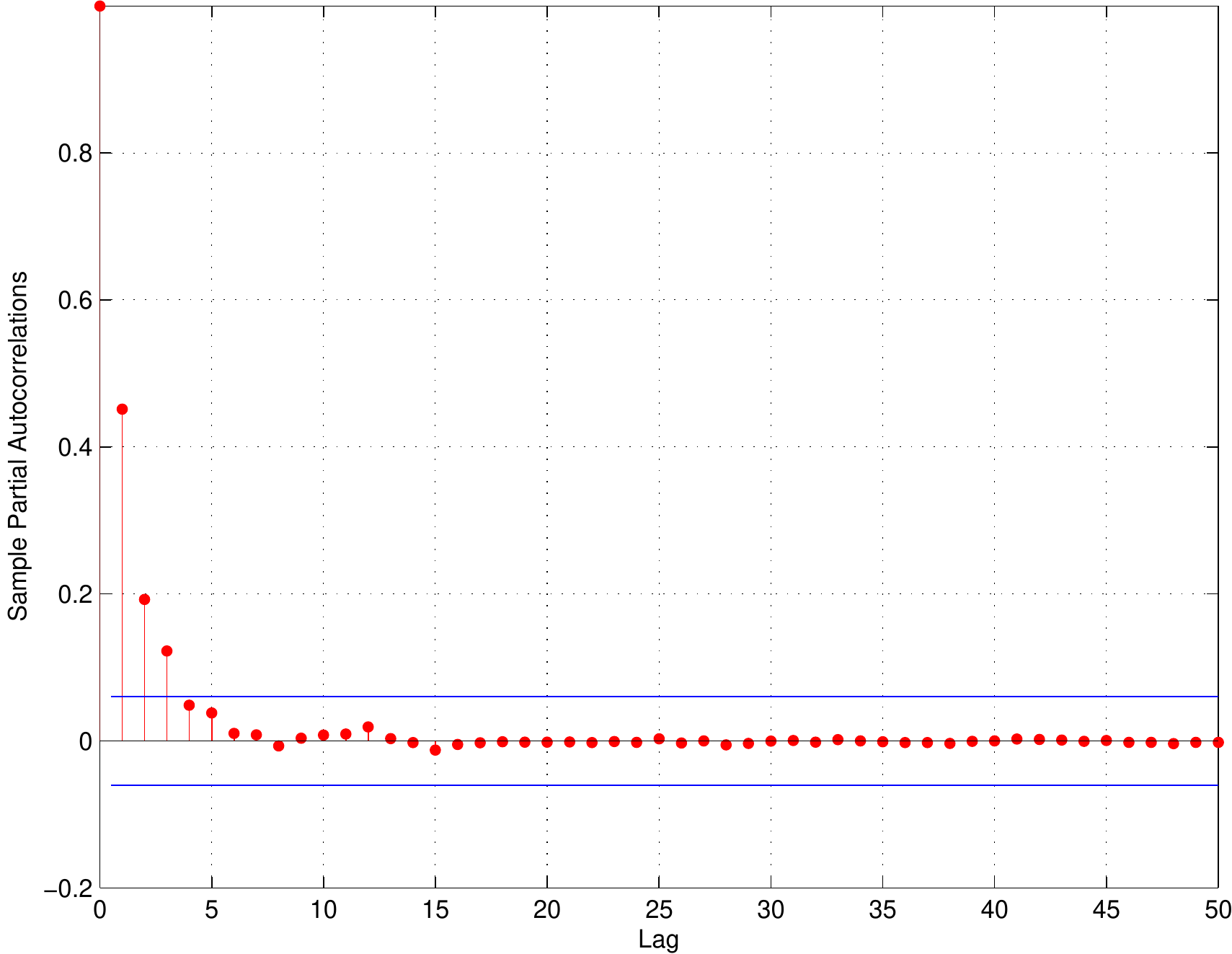}}
    \caption{Switching network type (`w') network example using 3 A-type memristors in circuit 7: a: $I-t$ curve in response to constant voltage; b: a periodogram; c: autocorrelation function; d: partial autocorrelation function. This demonstrates that increasing compositional complexity can move an ideal network away from ideality.}
    \label{fig:181012MTD6p1}
\end{figure}

\begin{figure}[!tbp]
    \centering
    \subfigure[]{\includegraphics[width=0.49\textwidth]{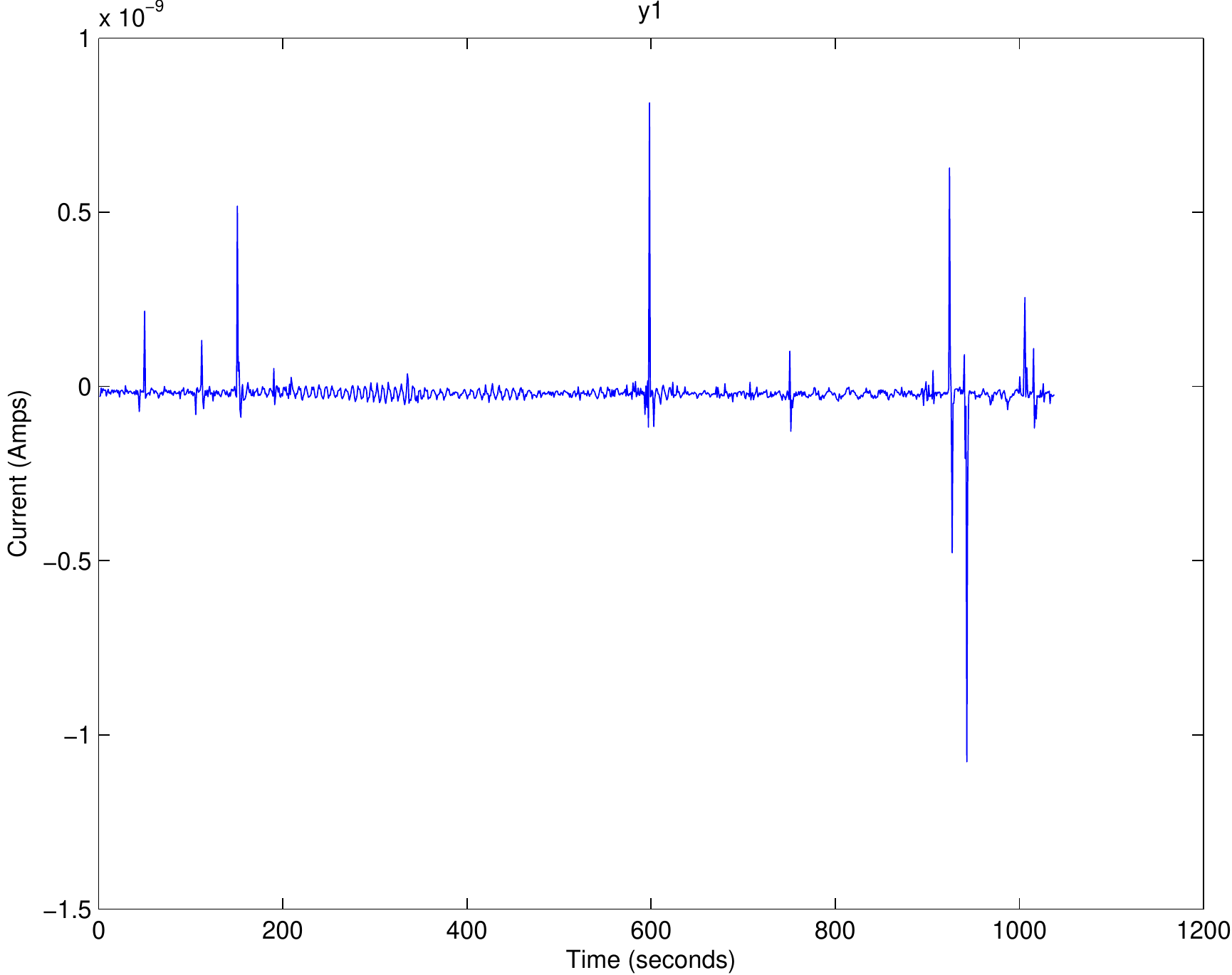}}
    \subfigure[]{\includegraphics[width=0.49\textwidth]{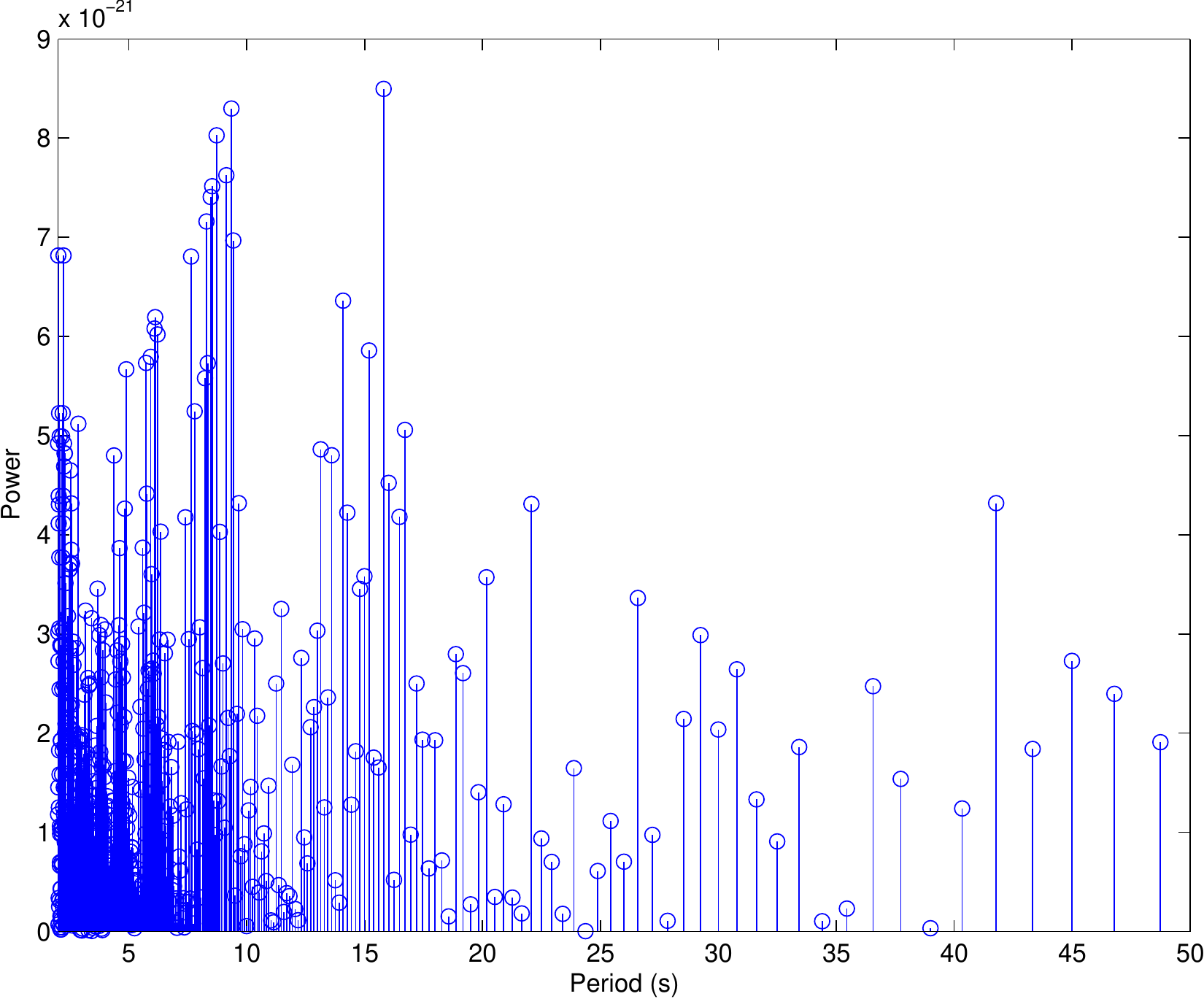}}
    \subfigure[]{\includegraphics[width=0.49\textwidth]{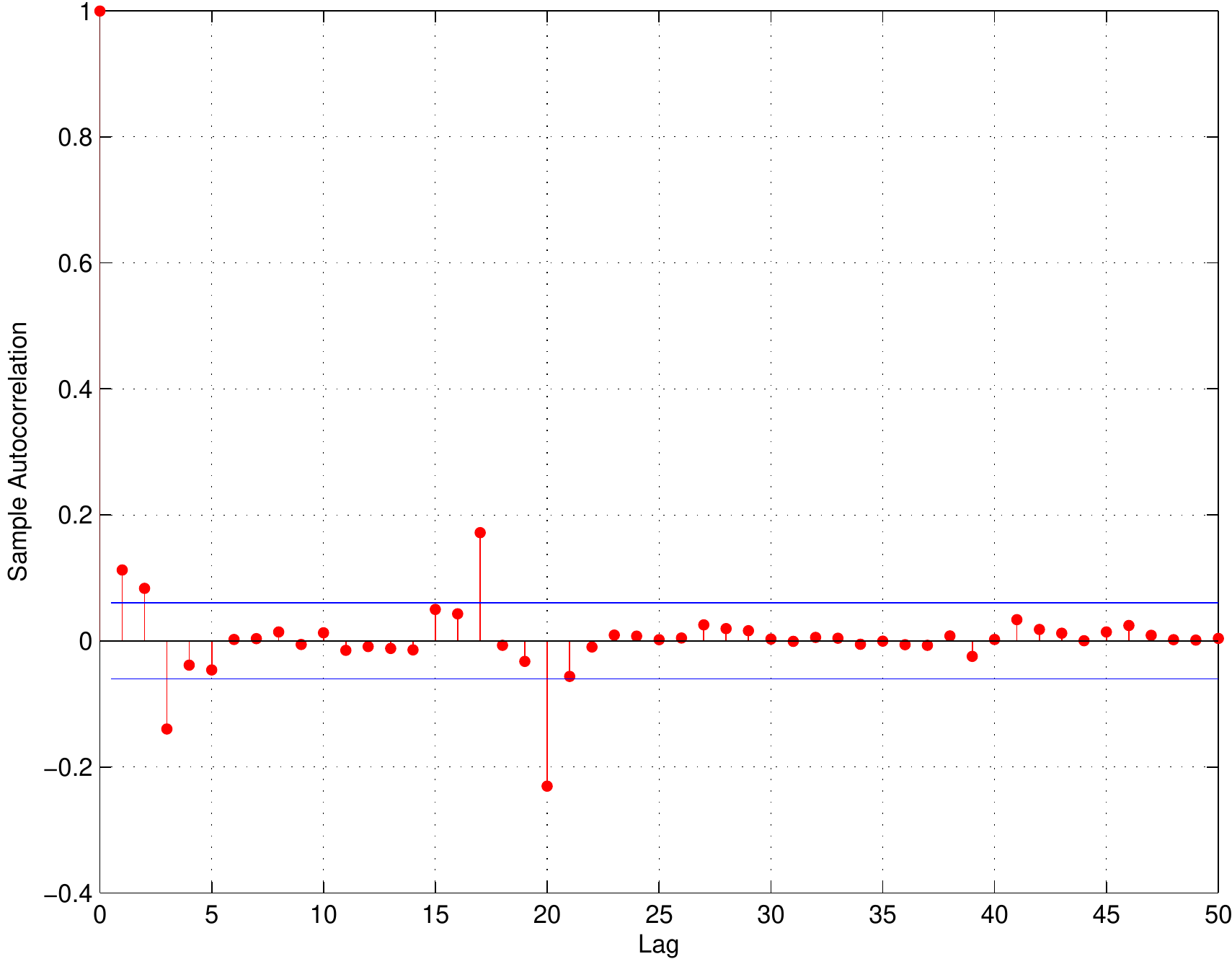}}
    \subfigure[]{\includegraphics[width=0.49\textwidth]{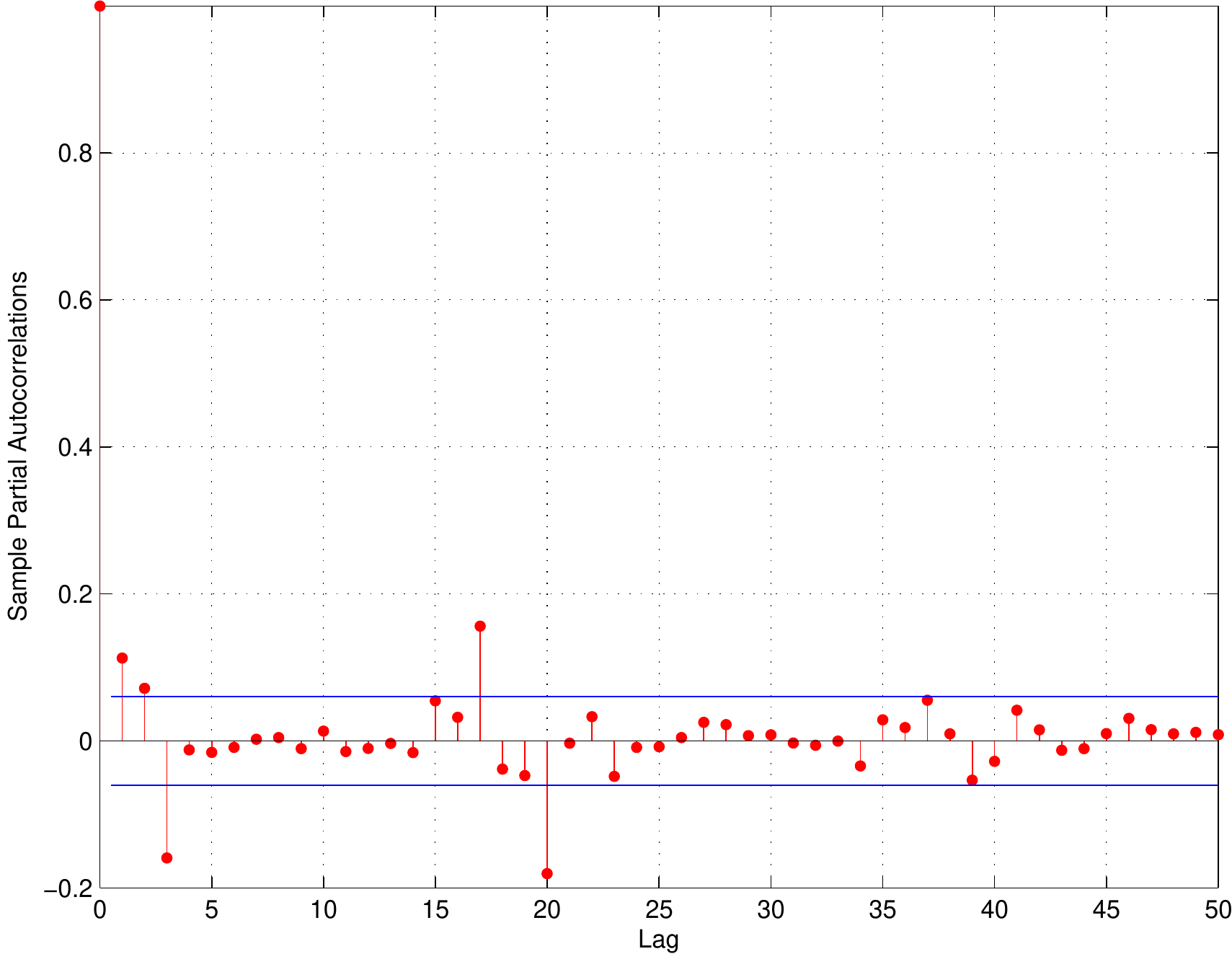}}
    \caption{Spiking network type `SpO' example from B type memristors in circuit 7: a: $I-t$ curve in response to constant voltage; b: a periodogram; c: autocorrelation function; d: partial autocorrelation function. The ACF clearly shows the interaction of several oscillations and a dampening directly after the single peak, the PACF shows we need an AR of order 7 to fit the data. }
    \label{fig:151012MT1p5}
\end{figure}

\begin{figure}[!tbp]
    \centering
    \subfigure[]{\includegraphics[width=0.49\textwidth]{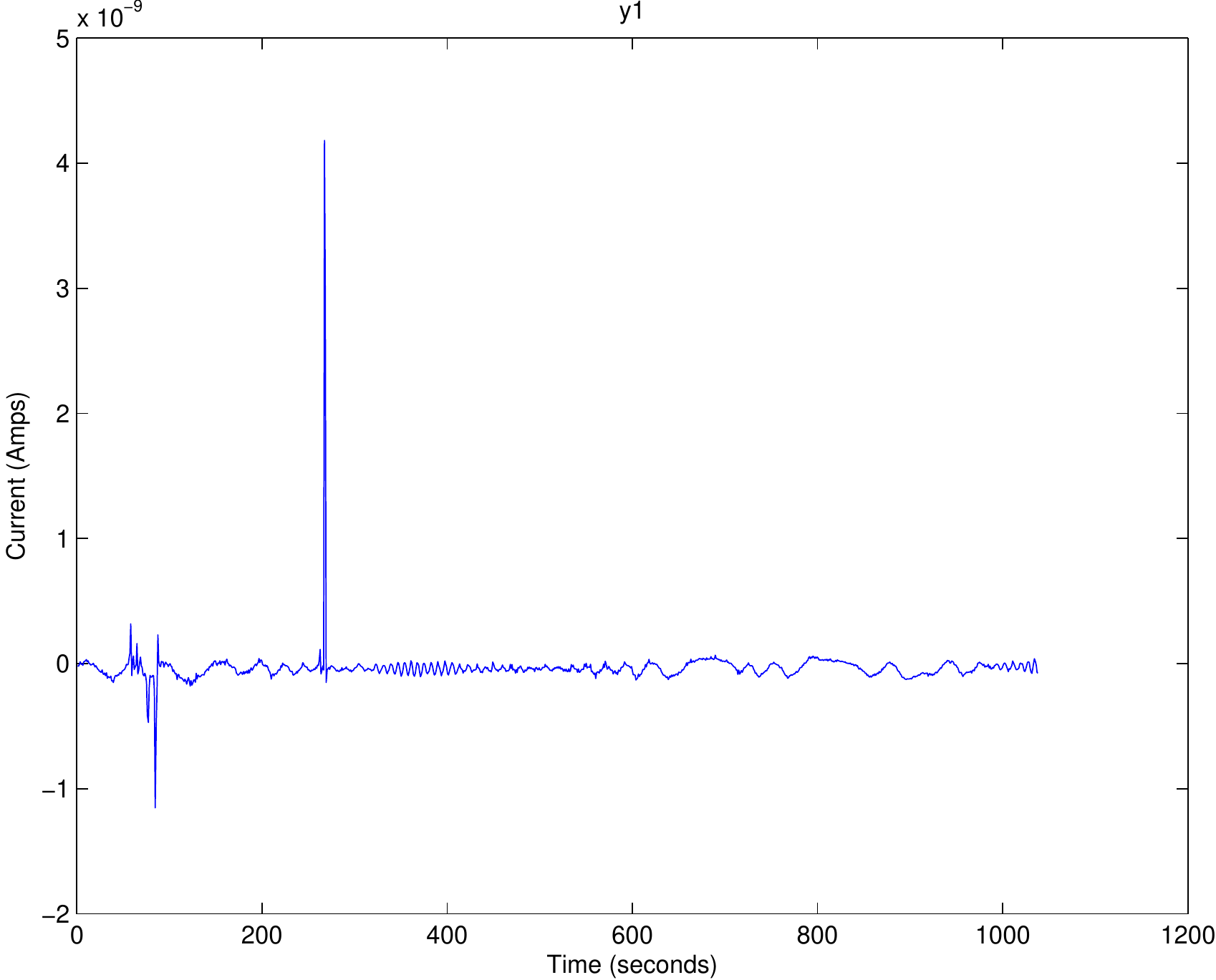}}
    \subfigure[]{\includegraphics[width=0.49\textwidth]{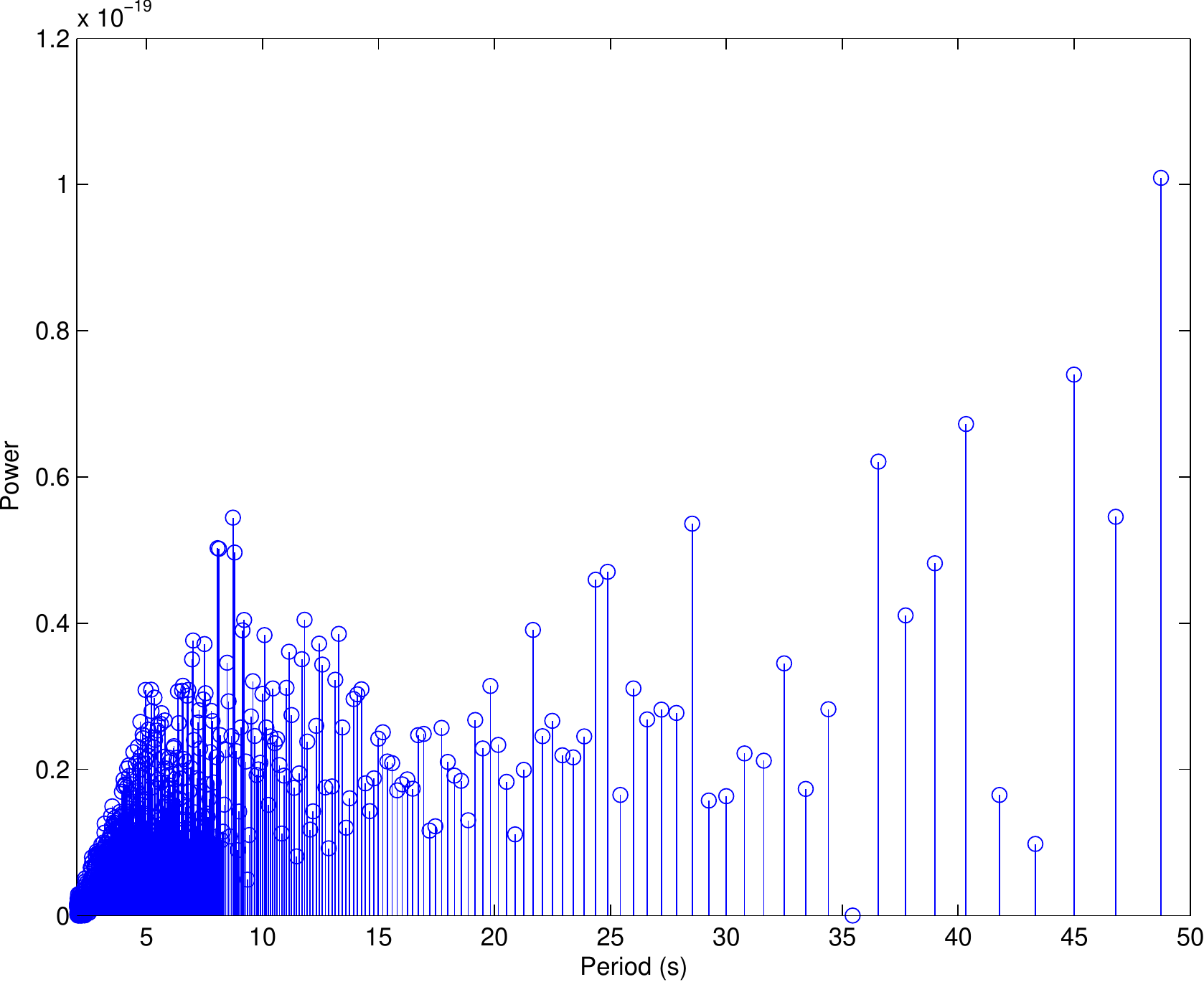}}
    \subfigure[]{\includegraphics[width=0.49\textwidth]{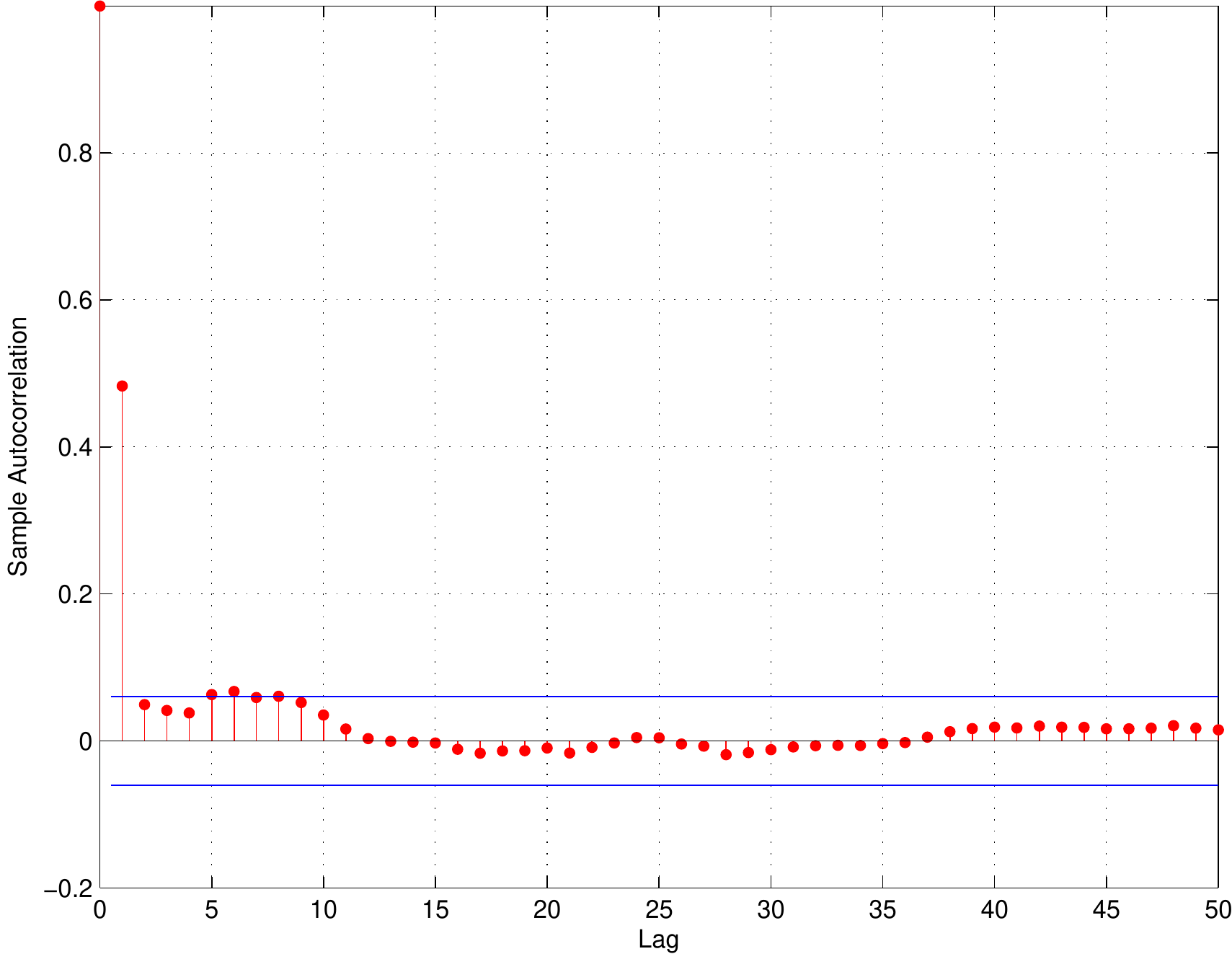}}
    \subfigure[]{\includegraphics[width=0.49\textwidth]{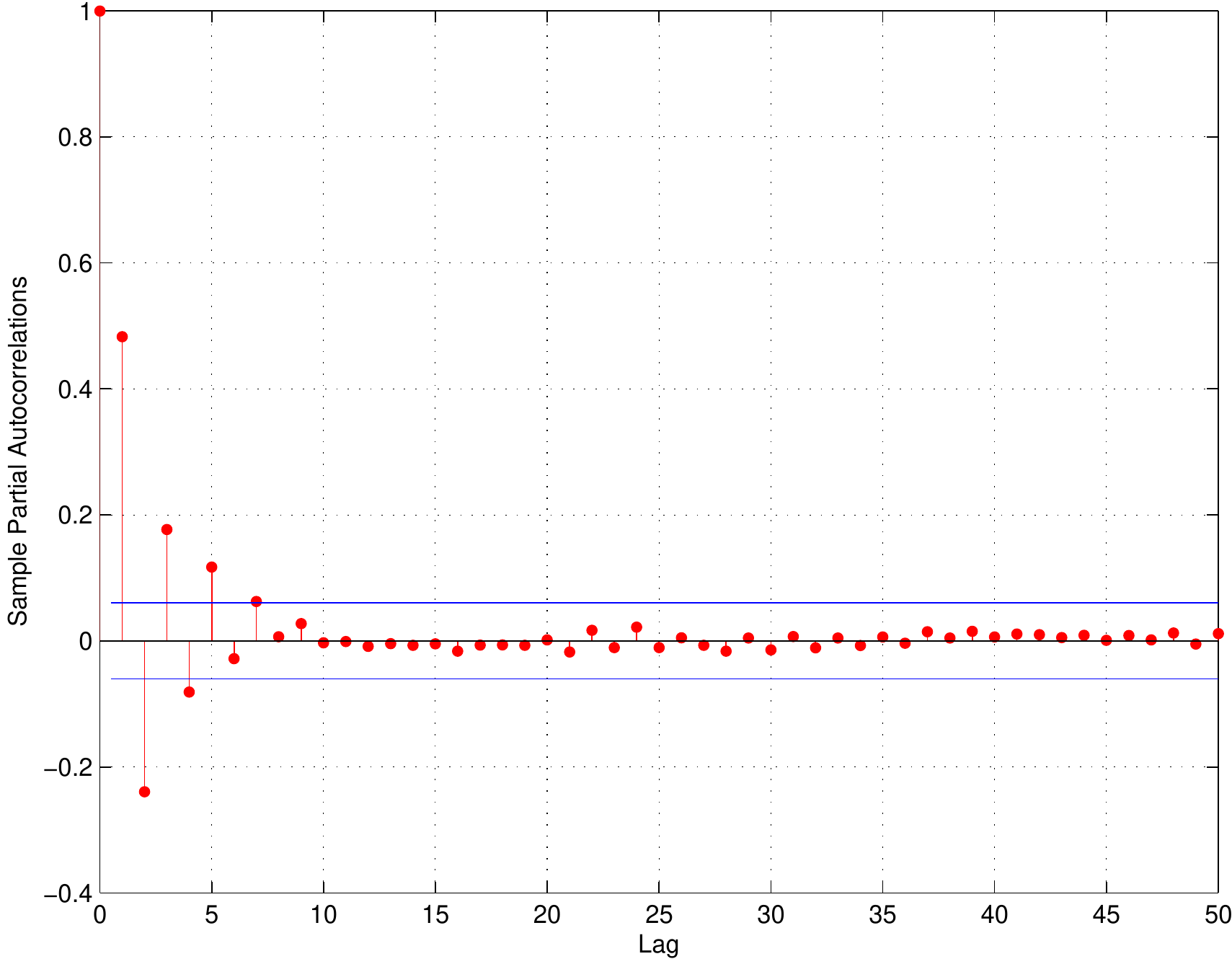}}
    \caption{Spiking network type `SpO' example for B-type memristors in circuit 5: a: $I-t$ curve in response to constant voltage; b: a periodogram; c: autocorrelation function; d: partial autocorrelation function. The ACF clearly shows the interaction of several oscillations and a dampening directly after the single peak, the PACF shows we need an AR of order 7 to fit the data. }
    \label{fig:241012MTR5p2}
\end{figure}

\begin{figure}[!tbp]
    \centering
    \subfigure[]{\includegraphics[width=0.49\textwidth]{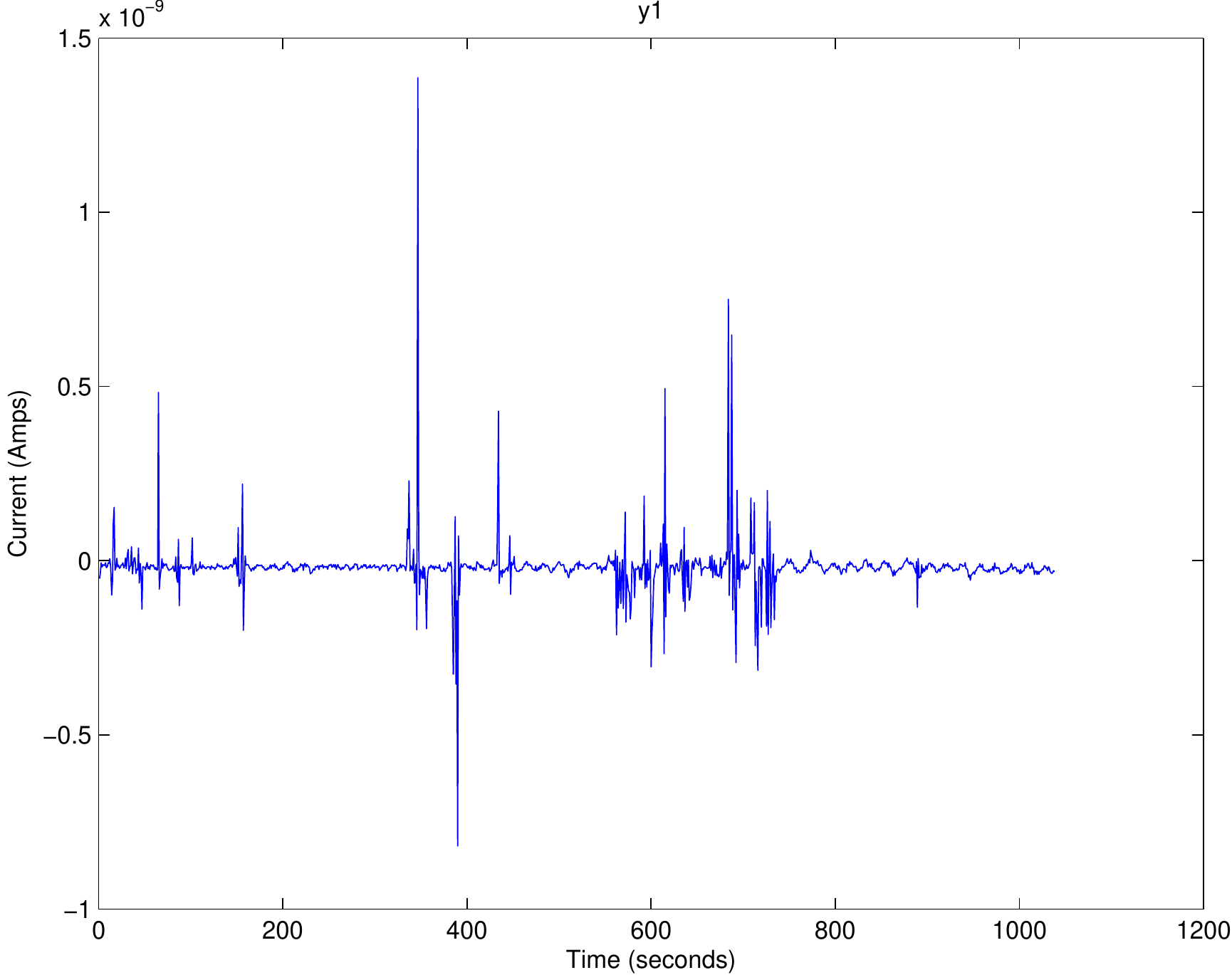}}
    \subfigure[]{\includegraphics[width=0.49\textwidth]{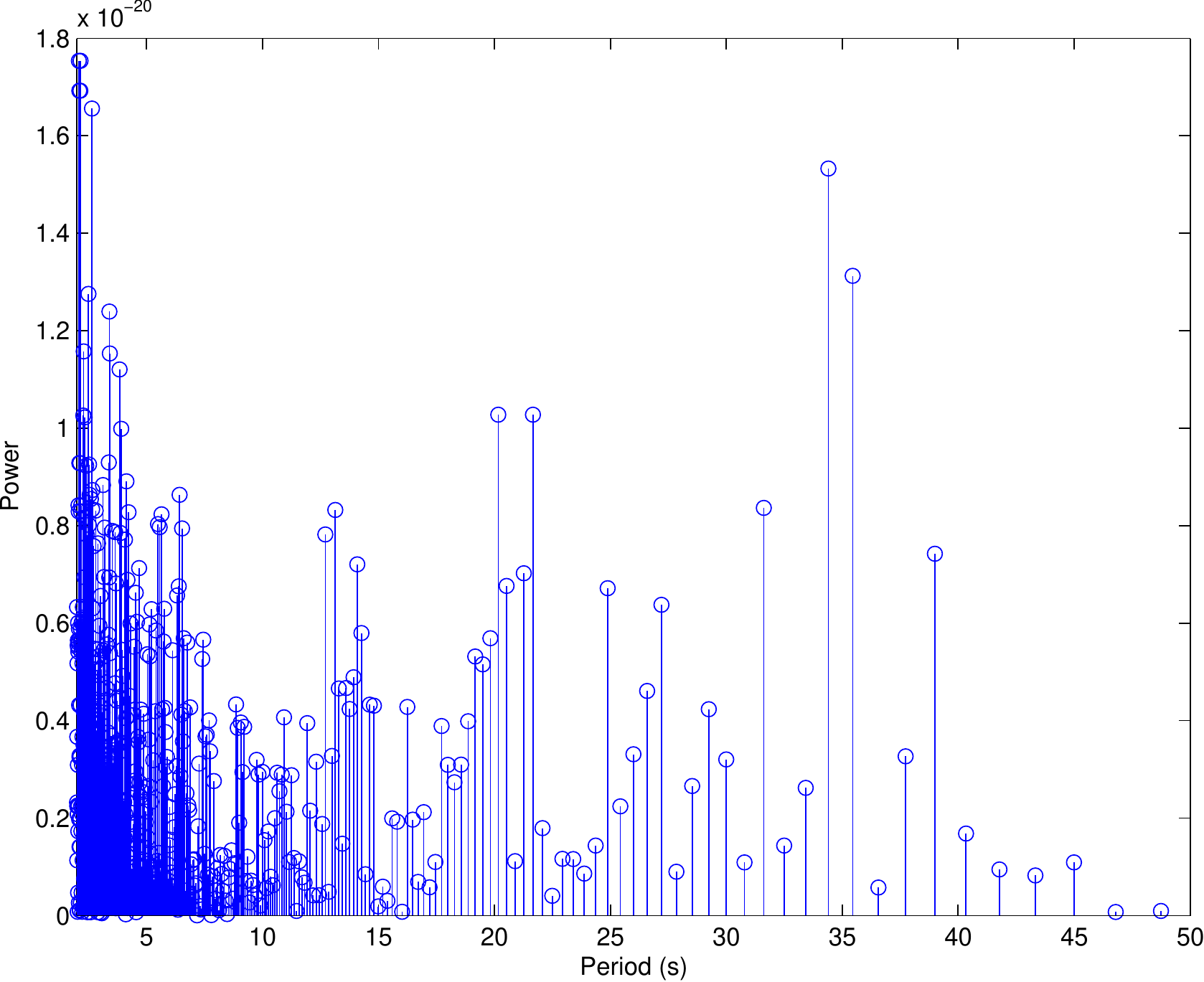}}
    \subfigure[]{\includegraphics[width=0.49\textwidth]{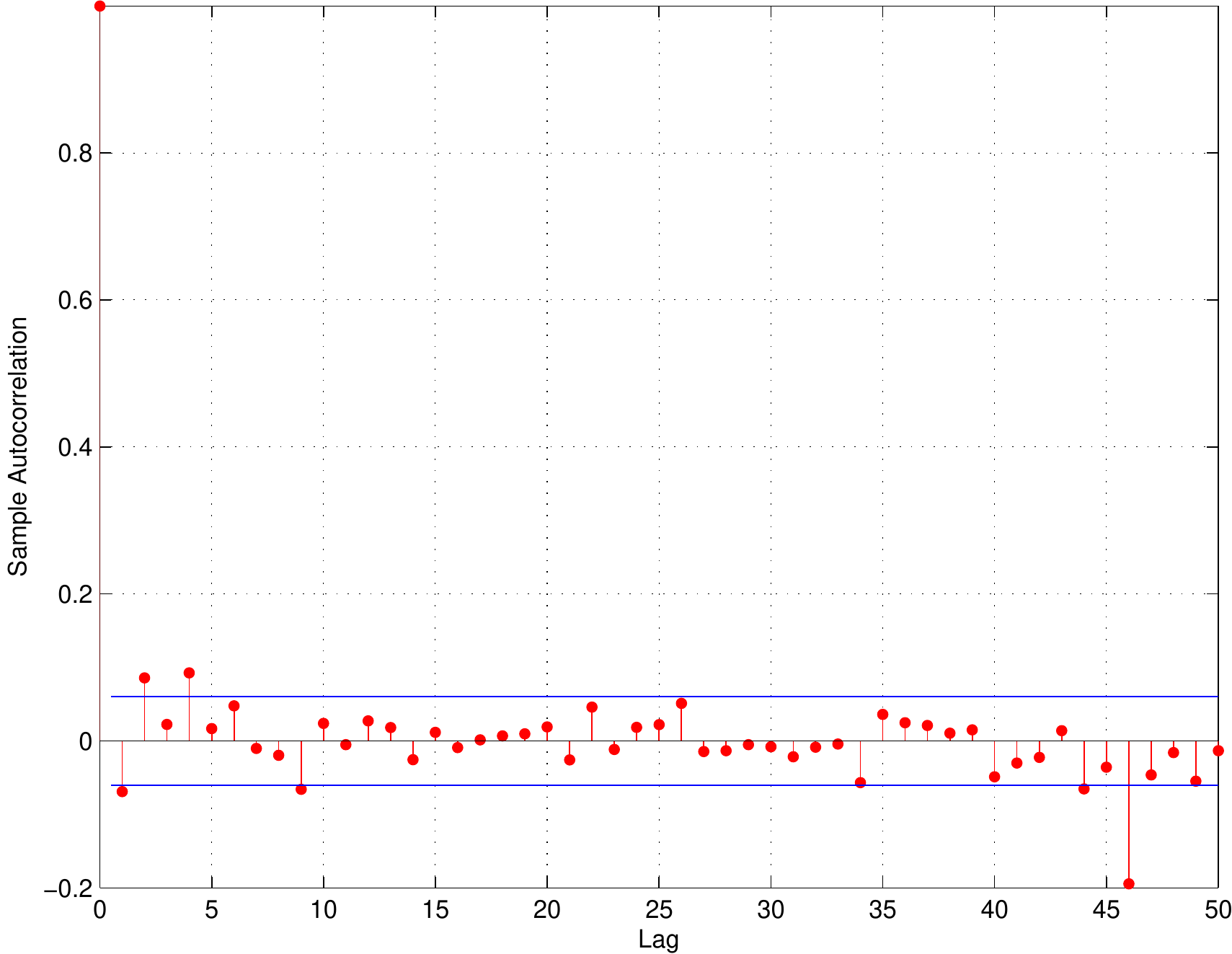}}
    \subfigure[]{\includegraphics[width=0.49\textwidth]{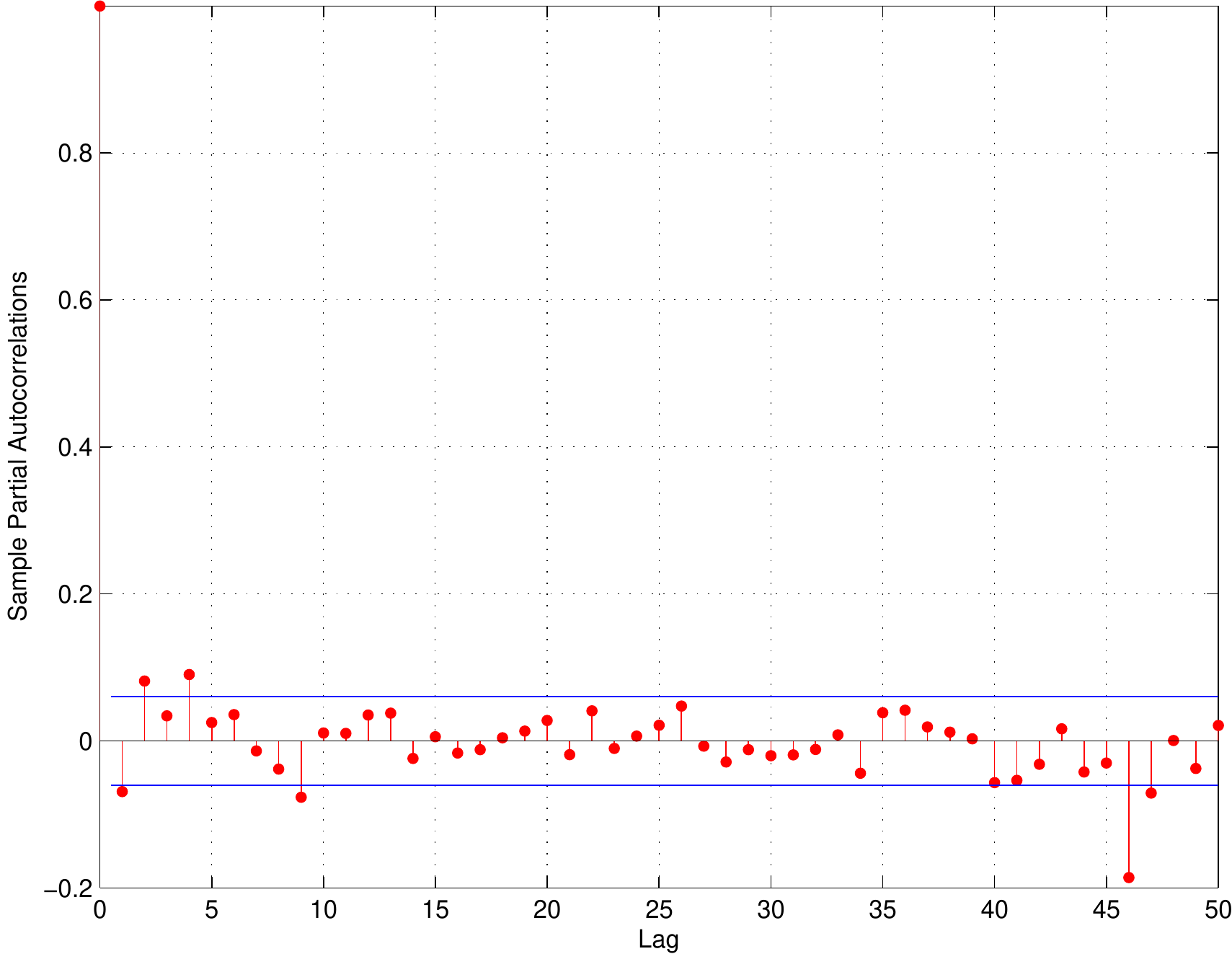}}
    \caption{Spiking network type `Sp' example of 3 B type memristors in circuit 7. : a: $I-t$ curve in response to constant voltage; b: a periodogram; c: autocorrelation function; d: partial autocorrelation function. We see groups of bursting spikes and underlying oscillations.}
    \label{fig:101012MT1p1}
\end{figure}

\begin{figure}[!tbp]
    \centering
    \subfigure[]{\includegraphics[width=0.49\textwidth]{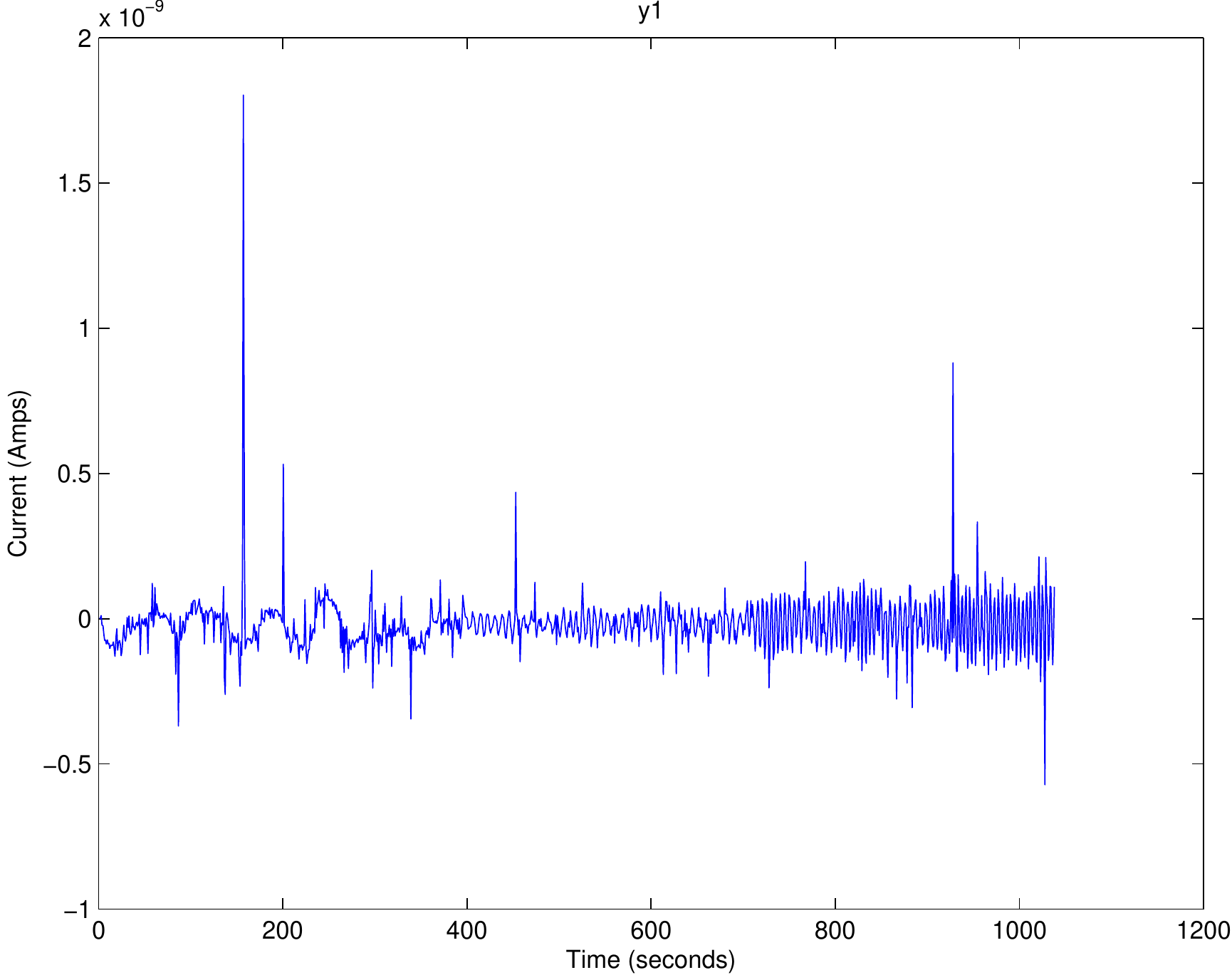}}
    \subfigure[]{\includegraphics[width=0.49\textwidth]{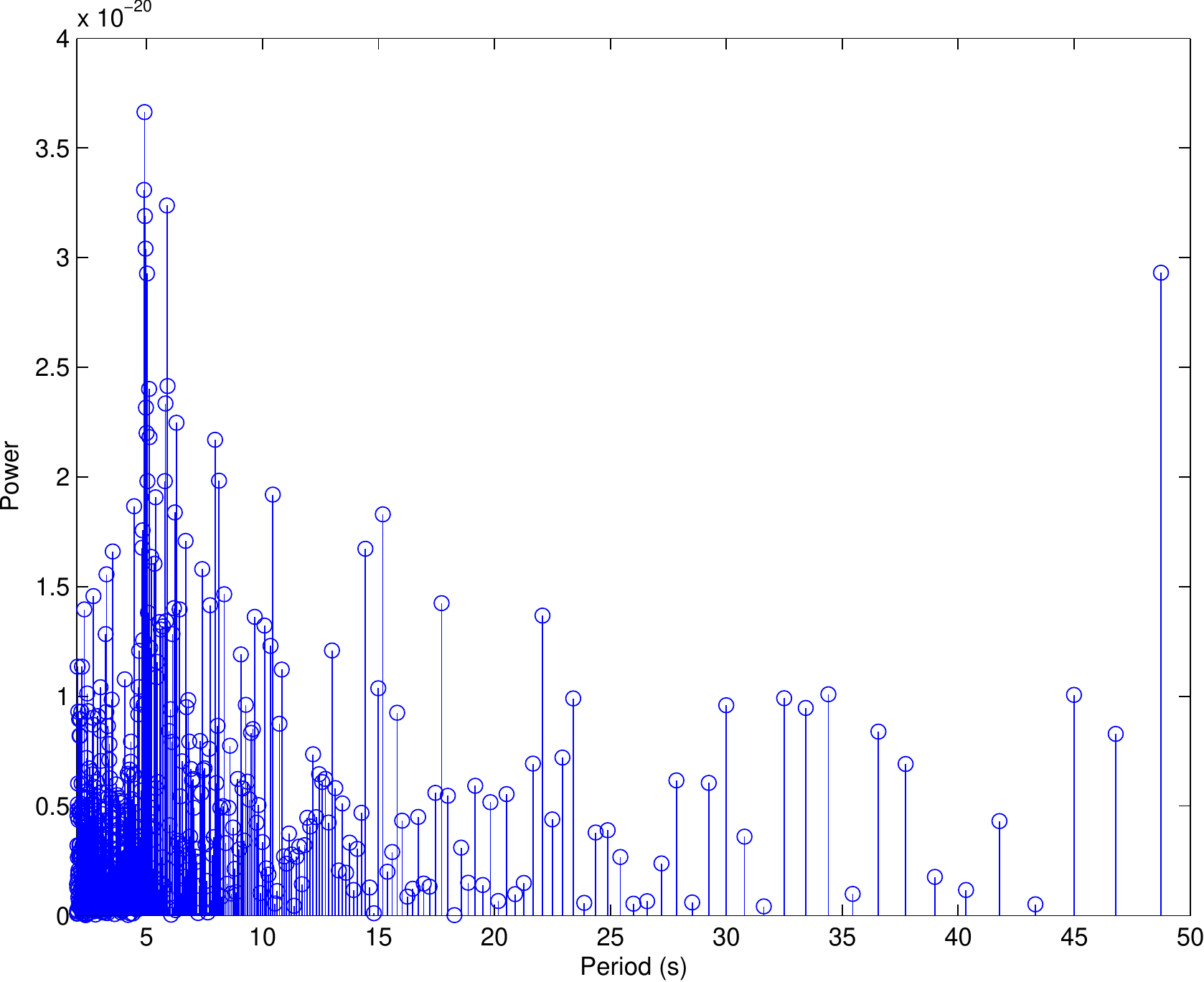}}
    \subfigure[]{\includegraphics[width=0.49\textwidth]{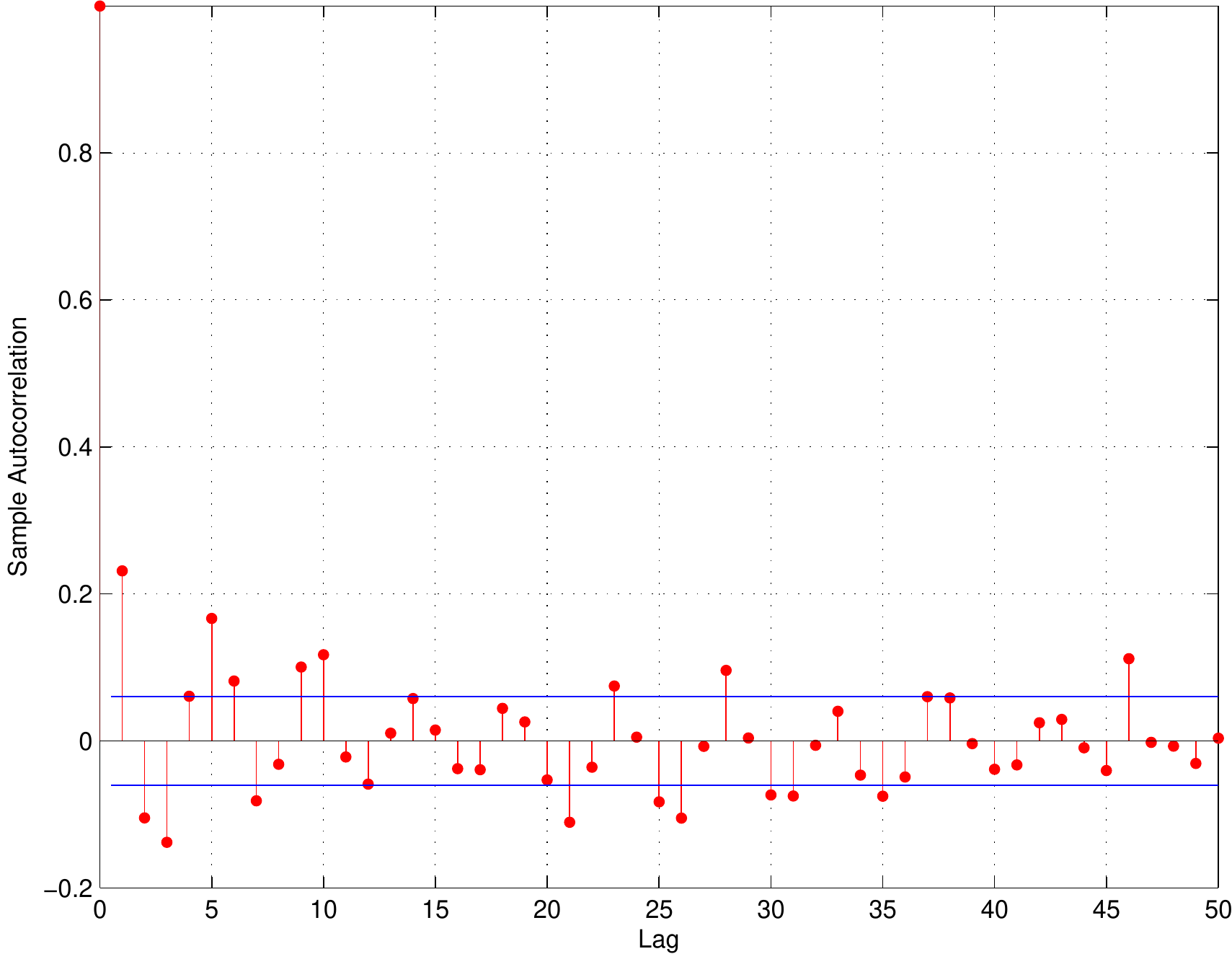}}
    \subfigure[]{\includegraphics[width=0.49\textwidth]{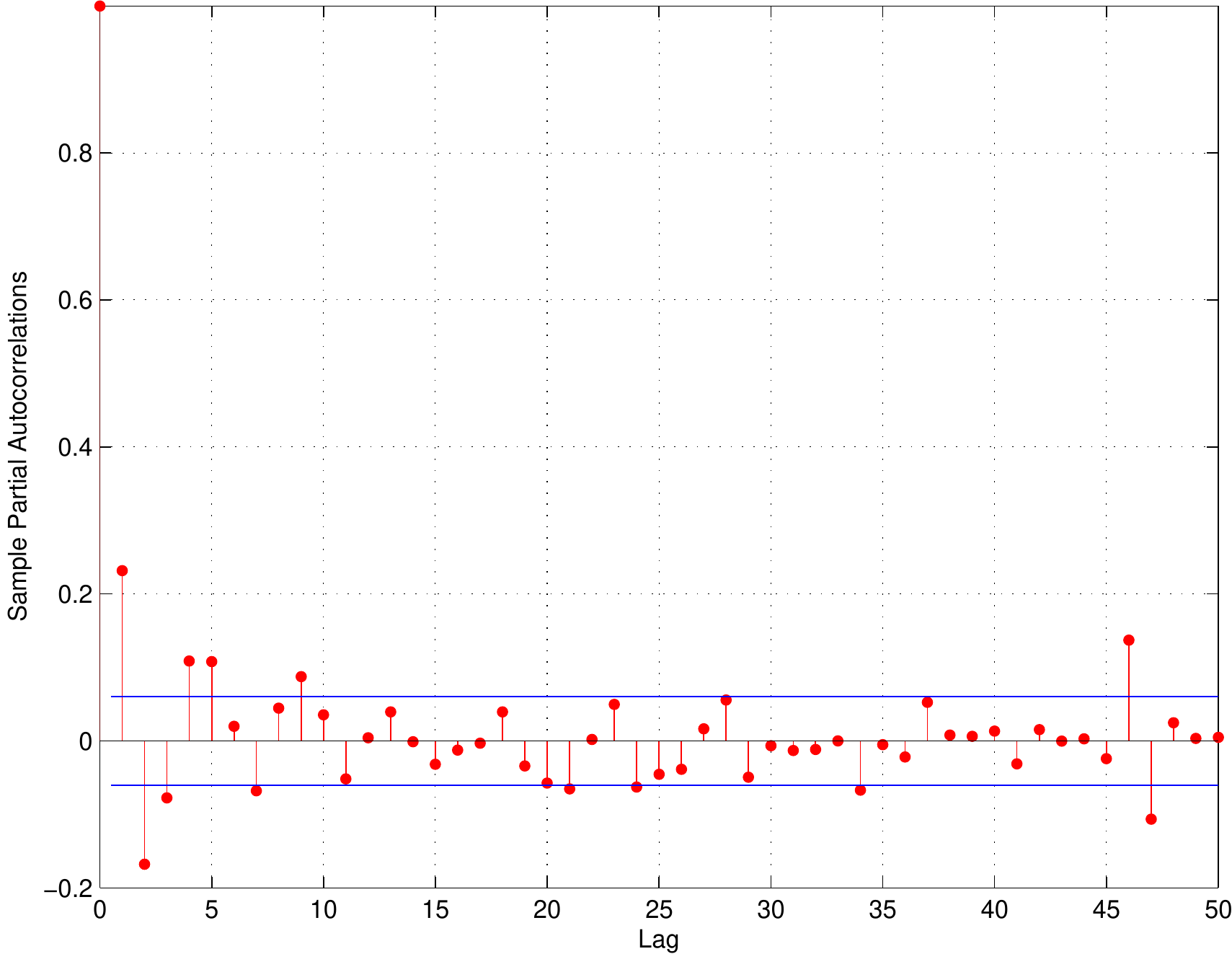}}
    \caption{An example of an oscillatory spiking `SpO' network: a: $I-t$ curve in response to constant voltage; b: a periodogram; c: autocorrelation function; d: partial autocorrelation function. The ACF shows the type of pattern expected for a small number of interacting sinusoids.}
    \label{fig:221012MTR6p3}
\end{figure}

\bibliographystyle{unsrt}

\bibliography{UWELit}




\end{document}